\documentclass[11pt]{iopart}
\usepackage{epsfig,here,hangcaption,rotating,amssymb}
\def\etal{{\it et al.}}

\def\gtap{\raisebox{-.4ex}{\rlap{$\sim$}} \raisebox{.4ex}{$>$}}

\def\lsp{{\tilde\chi_1^0}}
\def\ra{\rightarrow}

\def\tchi{\tilde\chi}

\let\badcite=\cite
\def\cite{~\badcite}

\newcount\timecount
\newcount\hours \newcount\minutes  \newcount\temp \newcount\pmhours
\hours = \time
\divide\hours by 60
\temp = \hours
\multiply\temp by 60
\minutes = \time
\advance\minutes by -\temp
\def\hour{\the\hours}
\def\minute{\ifnum\minutes<10 0\the\minutes
            \else\the\minutes\fi}
\def\clock{
\ifnum\hours=0 12:\minute\ AM
\else\ifnum\hours<12 \hour:\minute\ AM
      \else\ifnum\hours=12 12:\minute\ PM
            \else\ifnum\hours>12
                 \pmhours=\hours
                 \advance\pmhours by -12
                 \the\pmhours:\minute\ PM
                 \fi
            \fi
      \fi
\fi
}

\def\monthname{\relax\ifcase\month 0/\or January\or February\or
   March\or April\or May\or June\or July\or August\or September\or
   October\or November\or December\else\number\month/\fi}

\def\bold#1{\setbox0=\hbox{$#1$}%
     \kern-.025em\copy0\kern-\wd0
     \kern.05em\copy0\kern-\wd0
     \kern-.025em\raise.0433em\box0 }

\newcommand{\newc}{\newcommand}

\newc{\chioi}{\tilde{\chi}^0_1}
\newc{\chioii}{\tilde{\chi}^0_2}
\newc{\chipm}{\tilde{\chi}^{\pm}_1}
\newc{\staui}{\tilde{\tau}_1}
\newc{\slr}{\tilde{l}_R}
\newc{\sql}{\tilde{q}_L}
\newc{\sqr}{\tilde{q}_R}
\newc{\glss}{\tilde{g}}
\newc{\ohsq}{\Omega_{\chi}h^2}
\newc{\mhlf}{m_{1/2}}
\newc{\mo}{m_0}
\newc{\tgbet}{\tan \beta}
\newc{\etm}{E_T^{miss}}
\newc{\ao}{A_0}
\newc{\ie}{{\it i.e.}}
\newc{\eg}{{\it e.g.}}
\newc{\ssard} {{\tt SSARD}} 
\newc{\micromegas} {{\tt micrOMEGAs}}
\newc{\comphep} {{\tt CompHEP}}
\newc{\isasusy} {{\tt ISASUSY}} 
\newc{\isasugra} {{\tt ISASUGRA}} 
\newc{\isajet} {{\tt ISAJET}} 
\newc{\pythia} {{\tt PYTHIA}}

\def\beq{\begin{equation}}
\def\eeq{\end{equation}}

\def\ga{\mathrel{\raise.3ex\hbox{$>$\kern-.75em\lower1ex\hbox{$\sim$}}}}
\def\la{\mathrel{\raise.3ex\hbox{$<$\kern-.75em\lower1ex\hbox{$\sim$}}}}

\def\gtap{\raisebox{-.4ex}{\rlap{$\sim$}} \raisebox{.4ex}{$>$}}
\def\gev{{\rm \, Ge\kern-0.125em V}}
\def\tev{{\rm \, Te\kern-0.125em V}}
\def\gyr{{\rm \, G\kern-0.125em yr}}
\def\ohsq{\Omega_{\chi} h^2}

\def\gappeq{\mathrel{\rlap {\raise.5ex\hbox{$>$}}
{\lower.5ex\hbox{$\sim$}}}}
\def\lappeq{\mathrel{\rlap{\raise.5ex\hbox{$<$}}
{\lower.5ex\hbox{$\sim$}}}}
\def\Toprel#1\over#2{\mathrel{\mathop{#2}\limits^{#1}}}

\def\sl{{\widetilde \ell}_{\scriptscriptstyle\rm R}}

\def\m12{m_{1\!/2}}

\begin{document}
\topical[Cold Dark Matter and the LHC]{Cold Dark Matter and the LHC\footnote{This work was supported in part by the Director, 
 Office of Energy Research, Division of High Energy
Physics of the U.S. Department of Energy under Contract
DE-AC03-76SF00098}}
 \author[Marco~Battaglia,  Ian Hinchliffe, Daniel Tovey]{Marco~Battaglia$^{1,2}$,  Ian Hinchliffe$^2$, Daniel Tovey$^3$}
\address{$^1$University of California, Dept. of Physics, Berkeley, CA}
\address{$^2$Lawrence Berkeley National Laboratory, Berkeley, CA}
\address{$^3$University of Sheffield, Dept. of Physics and Astronomy,
    Hounsfield Road, Sheffield, S3 7RH, UK}


\begin{abstract}
{The recent determination of the dark matter density in the Universe by
the WMAP satellite has brought new attention to the interplay of
results from particle physics experiments at accelerators and from
cosmology. In this paper we discuss the prospects for finding direct
evidence for a candidate dark matter particle at the LHC and the
measurements which would be crucial for testing its compatibility with
cosmology data.}
\end{abstract}


\section{Introduction}

\label{sec:intro}

Recent precision data on the Cosmic Microwave Background (CMB) and
other astrophysical measurements have confirmed that a substantial
fraction of the mass of the universe is in the form of dark matter;
that is matter which which cannot be observed directly using
conventional astrophysical techniques. The COBE \cite{Wright:tf} satellite provided
evidence for intrinsic anisotropies in the CMB spectrum. The superior
resolution of the Wilkinson Microwave Anisotropy Probe (WMAP) mission
has enabled the accurate extraction of the baryonic and dark matter
densities of the universe. This has confirmed that the dark matter
density exceeds that of baryonic matter in the universe by a factor of
six~\cite{Spergel:2003cb}.  The mass density is expressed as a
fraction (referred to as $\Omega$) of the amount required for an
asymptotically flat universe and it is presently known to an accuracy
of 7~\%. The CMB results agree well with data obtained from supernovae
and the rotation curves of spiral galaxies.

Dark matter, as opposed to dark energy, can exist in two forms:
non-luminous baryonic matter in the form of large planets or dead
stars (MACHOS), or weakly interacting elementary particles that
pervade large regions of space. Searches for MACHOS have concluded
that these cannot be responsible for a substantial part of the dark
matter \cite{Alcock:1998fx}.

Particle physics candidates for dark matter fall into three basic
categories depending upon their masses and interactions. Weakly
interacting particles that are in thermal equilibrium at a very early
stage in the evolution of the universe will eventually fall out of
equilibrium as the universe expands.  The decoupling time (or
temperature) when this occurs depends on the expansion rate of the
universe as well as the couplings of these particles to other
particles that are still in equilibrium.  Particles that are
(non-)relativistic at the time that galaxies start to form are
referred to as (cold) hot dark matter. The simplest example of hot
dark matter  is a neutrino with a very small mass ($< {\cal{O}}(20)$~eV) and
of cold dark matter a very heavy neutrino ($\sim$~100~GeV). In both cases
the interaction rates are determined by the Standard Model of
electroweak interactions (SM).

The third type of particle dark matter can arise during the QCD phase
transition as the universe cools down. In this case the result can be
a gas of axions~\cite{Peccei:1977hh}.  For certain values of the mass
and coupling, these could account for a substantial part of the dark
matter~\cite{Asztalos:2001jk}.

Hot and cold dark matter have implications for the distribution of
visible matter (galaxies) throughout the universe~\cite{Davis:ui}.
The observed non-uniform distribution of such matter is believed to
arise from the growth of quantum fluctuations remaining in the early
universe after the end of inflation. These fluctuations are amplified
by gravitational interactions and lead to the clumping that we observe today. The dark
matter plays a vital role in this process as it dominates the total
matter. Hot dark matter is free streaming and produces a universe with
too little small scale structure. Cold dark matter is therefore
strongly favored~\cite{Spergel:2003cb}.

There are models of particle physics that provide natural candidates
for relic particles comprising the cold dark matter. This article will
discuss these candidates and how they could be produced and their
properties inferred from experiments at the LHC. We shall comment on
the interplay between the LHC experiments and terrestrial experiments
that aim to detect the dark matter directly.  Most of our discussion
focuses on supersymmetric candidates.

Many of the candidate dark matter particles can be produced copiously
at the LHC either directly or as decay products of the other
particles. Measurements at the LHC will then provide information
regarding the masses and couplings of both the dark matter candidate
and the other particles with which it interacts and which are
important in calculating its ``relic'' density.

In section \ref{sec:cdm-susy} we review methodology by which the relic
density is computed and discuss how the dark matter density constrains
the possible supersymmetric models. We then discuss the constrained
MSSM (CMSSM) model for which most of the detailed simulations have
been carried out. The consequences of relaxing some of the CMSSM
constraints are then discussed. This section concludes with a
discussion of dark matter candidates in gauge and anomaly mediated
models and of the possibility that sneutrinos or gluinos could account
for the dark matter.  Section~\ref{sec:lhc} then discusses the
potentially relevant LHC measurements of supersymmetric particles. The
possibility that particles arising from models of extra dimensions
could account for dark matter is discussed in
Section~\ref{sec:ued}. Section~\ref{sec:direct} then summarizes the
constraints that can be obtained from direct searches for dark
matter. In section~\ref{sec:complement} we illustrate how the
measurements from LHC, astrophysics and direct detection complement
each other and can be used to build up a complete picture of
supersymmetric dark matter.

\section{Cold Dark Matter and Supersymmetry}

\subsection{Supersymmetric Models}

\label{sec:cdm-susy}

Supersymmetry (SUSY) is a symmetry which relates fermions and
bosons. In supersymmetric theories all existing particles are
accompanied by partners having opposite spin-statistics. It is an
appealing concept, for which there is currently no direct experimental
evidence.  As none of the partner particles have been observed,
supersymmetry cannot be exact and must be broken so that the masses of
the superpartners are larger than a few hundred GeV
\cite{susy-limits}. Supersymmetry solves a long-standing problem with
the SM, namely that the electroweak scale ($\sim$ the mass
of the $Z$ and $W$ bosons) is of order 100~GeV even though radiative
corrections in the model point to a scale that is many orders of
magnitude larger, around the Planck scale ($10^{19}$~GeV ) where
gravity becomes strong.  A supersymmetric version of the SM is able to
stabilize these radiative corrections and solve this ``hierarchy
problem'' provided that the new particles have masses of less than a
few~TeV.

When supersymmetric models of electroweak interactions are
constructed, there often arises a symmetry, called R-parity, under
which all the existing particles are even and the new super-partners
are odd.  If R-parity is violated then supersymmetric models can
naturally give rise to unacceptably short nucleon lifetimes. If
R-parity is conserved, there are two important consequences; the
supersymmetric particles have to be produced in pairs and the lightest
of them is absolutely stable. In supersymmetric models the Lightest
Supersymmetric Particle (LSP) is invariably produced abundantly in the
early universe and hence if R-parity is conserved these LSPs should
still be present throughout the universe, providing a natural
candidate for the dark matter.

There are several candidates for the LSP, although many possibilities
are ruled out by rather general considerations. Most importantly the
LSP cannot carry electric charge as otherwise it would bind to
ordinary matter, violating the very stringent limits which exist on
exotic matter \cite{Smith:rz}. Remaining candidates are therefore the
gravitino (partner of the graviton), $\tilde{G}$, the sneutrino
(partner of the neutrino), $\tilde{\nu}$, the gluino (partner of the
gluon), $\tilde{g}$, and the lightest neutralino, $\tilde \chi^0_1$,
which is a mixture of the parters of the SM gauge bosons $Z^0$ and
$\gamma$ and the Higgs boson. In this last case, the coupling of the
LSP is determined by its content. It is most convenient to work in a
basis where the content is regarded as Higgsino, wino and bino, the
neutral fermion states of the supersymmetric Higgs sector and the
$SU(2)$ and $U(1)$ supersymmetric gauge theories. The first component
couples like a Higgs boson and therefore has very weak couplings to
quarks and leptons of the first two generations. The second component
couples more strongly than the third due to the greater $SU(2)$ gauge
coupling strength. 

The simplest SUSY model, known as the ``Minimal Supersymmetric
Standard Model'' or MSSM, is that containing one SUSY partner for each
of the known SM particles together with two Higgs doublets to generate
the masses of all the quarks and leptons. This model has over 100
parameters which control the masses and couplings of the new
supersymmetric particles. These are determined by the mechanism
responsible for SUSY breaking in the MSSM. There are several
candidates for the breaking mechanism which produce qualitatively
different mass spectra. In general it is assumed that there is an
additional ``hidden'' sector of the theory with particles of very
large mass where SUSY exists as a broken symmetry. This breaking is
communicated to the partners of the SM particles via a mediation
mechanism.

Since gravity is known to exist, it is a natural candidate for the
SUSY breaking mechanism and since it couples to all particles equally
it is expected that all the sleptons, sneutrinos (partners of the
charged leptons and neutrinos) and squarks (partners of quarks) will
have a common mass at the energy scale where the mediation
operates. Now the three couplings of the SM (the strong, weak and
electromagnetic couplings) are precisely measured in current
experiments, as are their evolutions with energy. If these couplings
are extrapolated to very high energy assuming the existence of only
the SM particles and their SUSY partners (with masses around 1 TeV)
then it is found that they unify to a common value at a scale slightly
below the Planck scale.  This coincidence can be explained by
postulating that the strong, weak and electromagnetic interactions are
unified at this scale giving equal gluino and gaugino masses. These
ideas are implemented in the SUGRA or constrained MSSM (CMSSM) model
\cite{sugra}. This has a total of five parameters: the common scalar
mass $m_0$, which determines the squark, slepton, sneutrino and Higgs
masses, the common gaugino mass $m_{1/2}$, which determines the
gaugino and gluino masses, $\tan\beta$ which controls the relative
size of the couplings of the Higgs boson to up and down type quarks,
the sign of the Higgsino mass parameter $\mu$, and the common
trilinear coupling $A_0$, whose value is  relevant only in a few
cases. It is a remarkable feature of this model that,
as the parameters are evolved down to lower energies, the masses
change in a calculable way and, due to the large coupling of the top
quark to the Higgs, electroweak symmetry is broken spontaneously and
masses for the $W^{\pm}$ and $Z^0$ bosons generated automatically. As
this model is simple, and fully described by a small number of
parameters, it has been used most often for detailed phenomenological
studies. Other models of SUSY breaking have different structures and
we will comment on these later.

\subsection{Computing the Cold Dark Matter Density}

\label{sec:comp}

The maintenance of thermal equilibrium of particle densities in the
early universe depends upon the interaction rates of the particles and
the expansion rate of the universe. At very early times when the
universe is at a very high temperature, all particles are in thermal
equilibrium. Consider a particle that can annihilate against its
anti-particle with a cross-section $\sigma $.  The products of this
annihilation are other (lighter) particles. The production and
annihilation of the particles maintains their equilibrium. The time
evolution of the number density $n$ is described by:
\begin{equation}
\frac{dn}{dt} = - 3 H n - <\sigma v> (n^2 - n^2_{eq})
\end{equation} 
The first term on the right represents the dilution caused by the
expansion of the universe, with $H$ being the Hubble constant. The
second term represents the effect of the annihilation, with $v$ being
the velocity of the particles and $<\sigma v>$ representing a thermal
average of the cross-section times velocity. $n_{eq}$ is the number
density at thermal equilibrium given by a Boltzmann distribution. If
$\sigma$ is sufficiently large, the second term ensures that $n =
n_{eq}$, and hence the relic density is exponentially suppressed at
temperatures where the particles are non-relativistic. If $\sigma$ is
very small then the particles cannot remain in equilibrium. For most
processes interactions at sufficiently early time, or equivalently
high temperature, are dominant, and equilibrium is maintained until
the temperature falls below a freeze out temperature ($T_F$) when the
interaction rate becomes too low and annihilation ceases. The number
density at the present day is then given by $n_{eq}(T_F)$ diluted by
the Hubble expansion. The surviving number is most sensitive to
$\sigma$; a larger $\sigma$ ensures a lower number of survivors and a
smaller contribution to the mass density $\Omega$. As the particle's
mass increases its contribution to $\Omega$ also increases and
therefore $\sigma$ must be raised to satisfy experimental
constraints. We shall frequently refer to this contribution to
$\Omega$ arising from a specific dark matter particle as
$\Omega_{\chi} h^2$, where the normalisation is appropriate for
comparison with CMB data.

Amazingly, if supersymmetric particles have masses of order 100~GeV as
indicated by the fine tuning argument discussed above, their present
abundance is of the right magnitude to account for the ``observed'' cold
dark matter. It is this fact that has led to such a large focus on
supersymmetric dark matter candidates.  In practice, the computation
of the relic density of LSPs is more complicated than indicated
above. There can be several supersymmetric particles with comparable
masses, all of which are falling out of thermal equilibrium at the
same time. As well as particle-anti-particle annihilation, processes
involving the scattering of different species are important
(``co-annihilation'').

Several software packages have been developed to compute the cold dark
matter density in a supersymmetric model with conserved
R-parity~\cite{Belanger:2001fz,Gondolo:2002tz}. It is important to
emphasize that the masses and couplings of the supersymmetric
particles must be known before a prediction of the relic density can
be obtained. Renormalization group equations (RGEs) are used to evolve
the high scale CMSSM parameters and determine the soft parameters
(masses and couplings) at the electroweak scale. Even small
differences in these masses and couplings can result in significant
changes in the predicted relic density. A recent survey found that the
uncertainties in the masses of SUSY particles, defined by the
differences between predictions from four independent programs, can be
of order 20-30~\%~\cite{Allanach:2003jw}. Both {\tt
DarkSusy}~\cite{Gondolo:2002tz} {\tt v.~4.0} and
\micromegas~\cite{Belanger:2001fz} {\tt v.~1.3} can be interfaced to
several different codes for computing the full SUSY spectrum including
{\tt ISASUSY} \cite{Paige:2003mg} and {\tt SUSPECT}
\cite{Djouadi:2002ze}.

The main quantities which affect the relic density are the mass of the
LSP and its interactions, determined by its composition. Over much of
the CMSSM parameter space the LSP is the lightest neutralino, which
usually consists predominantly of the partner of the gauge boson of
the $U(1)$ sector of the SM (the bino). The coupling constant of this
is the smallest of the three SM couplings and the annihilation
cross-section is therefore small. The dominant mechanism contributing
to $\sigma$ is often $\lsp \lsp\to \ell^+\ell^-$ which includes
slepton exchange and is sensitive to the mass of the lightest slepton.
In the CMSSM therefore, as $m_0$ increases, $\sigma$ decreases and the
relic density increases. Generically, supersymmetric particle masses
cannot be too large if the universe is not to be overclosed by an
excess of dark matter ($\Omega > 1$).

In order to calculate the LSP relic density for all possible masses,
one must consider a very large number of possible SUSY annihilation
channels. In practice these contributions must be computed rapidly in
order to explore wide regions of the multi-dimensional supersymmetric
parameter space. An interesting solution, adopted by \micromegas~has
been to rely on the \comphep~program~\cite{Pukhov:1999gg} for
computing the cross-sections. Loop corrections can be important in
many cases, in particular those affecting widths of particles such as
the Higgs. Use of one-loop results rather than a tree level treatment
of the Higgs width can change the $\Omega_{\chi} h^2$ relic density
value by a factor of two at large $\tan \beta$, due to the enhancement
of the $b \bar b$ coupling.  Thanks to efforts over more than a
decade, the relic density can now be estimated from the electroweak
scale MSSM parameters with an accuracy which in principle approaches
1\%.  However, supersymmetric threshold corrections will depend on the
accuracy in determining the relevant SUSY particle masses. Also, the
full QCD and the higher-order electroweak corrections need to be
computed. If these are available by the time the LHC has collected a
significant data set, the intrinsic accuracy on the dark matter
density extracted from the supersymmetric particle spectrum should be
comparable to that expected from the next generation of
satellite-borne CMB experiments.

As the SUSY particle masses increase, cosmologically viable solutions
are possible only if the (co-)annihilation rates can be enhanced, for
instance by accidental mass degeneracies.  Furthermore, $\tchi^0_1$
annihilation through resonances may take place at large velocities,
thus requiring a relativistic treatment. A formalism for relativistic
thermal averaging was developed in Ref.~\cite{Edsjo:1997bg} and has
been adopted by several programs.

If the LHC experiments are able to determine the masses and couplings
of the SUSY particles, then the cold dark matter density can in
principle be predicted using these software packages. A small change
in the mass of the LSP may not have a significant effect on the LHC
phenomenology but can have a dramatic effect on the cold dark matter
density. Conversely, the masses of heavier SUSY particles such as
squarks and gluinos are very important at the LHC, as these are
usually the dominant source of produced SUSY particles, while their
masses are usually irrelevant to calculation of the relic density. A
consistent attempt to reconcile LHC data with that from dark matter
observations will require that the same model can be applied to
both.

\subsection{Cold Dark Matter in the Constrained MSSM}

\label{sec:cmssm}

\begin{figure}
\begin{center}
\epsfig{file=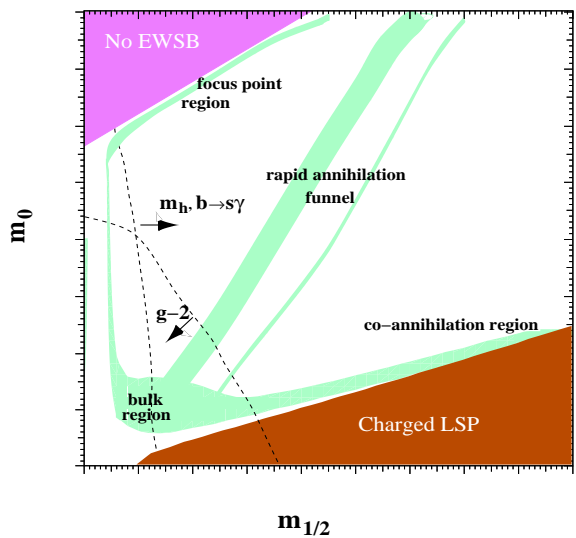,width=7.5cm,height=7.0cm} 
\end{center}
\caption{Schematic plot of the $m_0$-$m_{1/2}$ plane showing the
  regions where cosmologically interesting cold dark matter relic
  densities occur (light shaded regions). The dark region at low $m_0$
  is excluded by the requirement that the LSP be neutral (here it is a
  charged slepton), while the dark region at large $m_0$ is excluded
  by the requirement that electroweak symmetry breaking occur. If
  SUSY is to explain the possible anomaly in the measurement of the
  muon anomalous magnetic moment\cite{Bennett:2004pv}, the region to
  the left of the $g-2$ line is preferred.  A region at small
  $m_{1/2}$ is excluded by the direct limits on the Higgs boson mass
  \cite{Barate:2003sz}. The contribution from SUSY to the observed
  \cite{Chen:2001fj} decay $b \to s \gamma$ excludes the region at
  small masses unless the SUSY contribution is cancelled by other new
  effects.
\label{parplane}}
\end{figure}

Within the constrained MSSM class of models the LSP is almost always
the lightest neutralino. Cosmologically interesting LSP relic
densities mostly occur only in certain regions of $m_0$-$m_{1/2}$
parameter space illustrated schematically in
Figure~\ref{parplane}. The region at low $m_0$ and $m_{1/2}$ is
referred to as the ``Bulk Region''. Here there are no accidental
degeneracies of the SUSY particles, no enhancements of $\sigma$ and
the LSP has a mass of less than 200~GeV. The region extending from the
bulk region to large $m_{1/2}$, running along the edge where the
slepton becomes the LSP has an enhanced annihilation rate as the
lightest slepton and LSP are almost mass degenerate. This is known as
the ``co-annihilation tail''. The region at large values of $m_0$ and
$m_{1/2}$ is sometimes known as the ``rapid annihilation funnel'' or
``Higgs pole region'' and occurs when the mass of the LSP is such that
annihilation via an intermediate (s-channel) heavy Higgs boson ($A$)
enhances $\sigma$. Finally, at large $m_0$ there is the ``focus point
region'' or ``hyperbolic region'' along the boundary beyond which
electroweak symmetry breaking no longer occurs. We will now discuss in
more detail the phenomenology of each of these regions. Note however
that cosmologically interesting relic densities can also be generated
in certain other scenarios, for instance when the LSP is nearly mass
degenerate with the lighter stop squark at large $A_0$.

\subsubsection{Bulk Region and Co-Annihilation Tail}

\label{sec:cmssm1}

The region at low values of $m_0$ and $m_{1/2}$ corresponds to the
largest area of the CMSSM parameter space yielding a cosmologically
interesting cold dark matter relic density. In this region, the SUSY
particles are relatively light and therefore the region is severely
constrained by negative results of searches for supersymmetric
particles at LEP-2 and the Tevatron. In the CMSSM, the Higgs mass is
constrained to be less than 134~GeV \cite{higgsmasslimit}. Searches
for the Higgs boson at LEP result in a lower limit on the Higgs mass
of 114.4~GeV\cite{Barate:2003sz}. This and the lower limit on the
chargino mass \cite{susy-limits} reduce significantly the allowed
portion of this bulk region. Virtual effects from SUSY particles can
affect the rates for certain rare processes.  In particular, the
observation of the loop-mediated $b \to s \gamma$ process provides a
constraint which excludes small masses  and is particularly important at negative $\mu$
and large values of $\tan \beta$. If
SUSY is to explain the possible anomaly in the measurement of the muon
anomalous magnetic moment\cite{Bennett:2004pv}, small masses are
preferred.

\begin{figure}
\begin{center}
\begin{tabular}{c c}
\epsfig{file=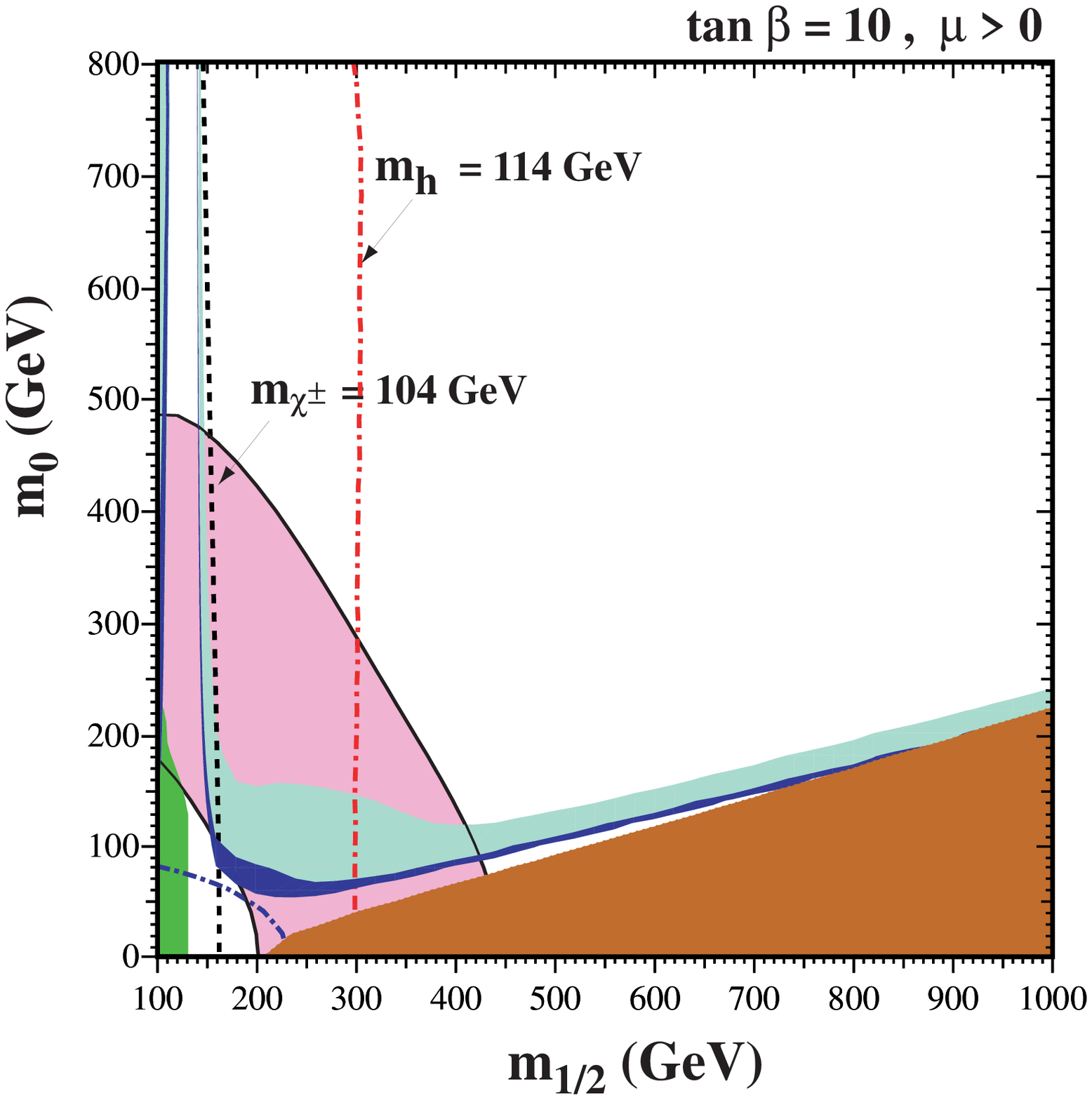,width=7.5cm,height=7.0cm} &
\epsfig{file=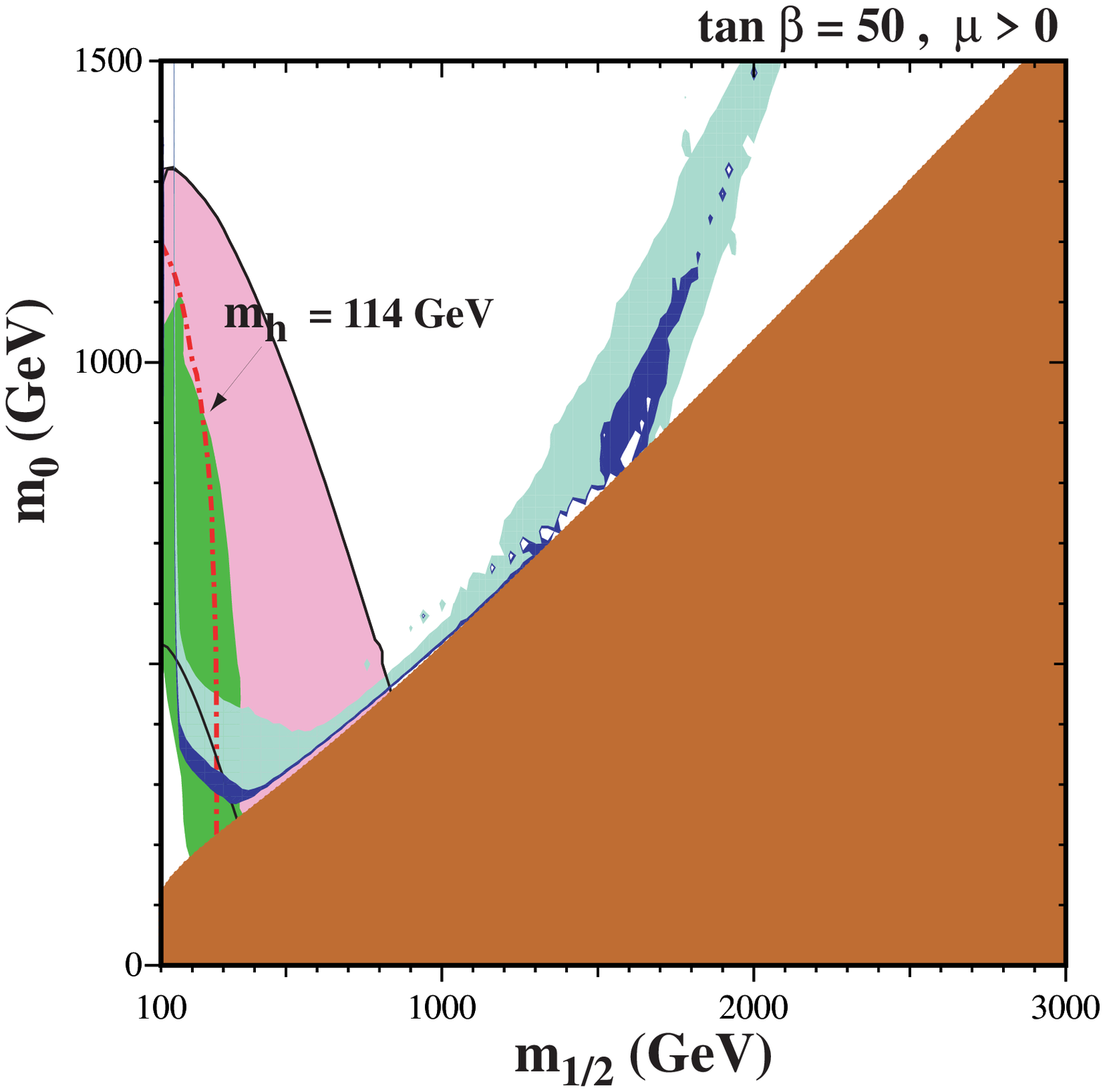,width=7.5cm,height=7.0cm} \\
\end{tabular}
\end{center}
\caption{The $m_0$-$m_{1/2}$ plane for $\tan \beta$=10 (left) and 50
(right) showing the region allowed by WMAP data and the existing
constraints from accelerator data. The ``L''-shaped dark (blue) and
light (cyan) regions correspond to $0.094 < \Omega_\chi h^2 < 0.129$
(WMAP) and $0.10 < \Omega_\chi h^2 < 0.30$ respectively.  The regions
on the left of the vertical lines are excluded by the LEP-2 limits on
the chargino (dashed) and the Higgs (dashed-dotted) masses.  The
region below the lower (brown) triangle is forbidden since here the
LSP is charged. From~Ref.\cite{Ellis:2003cw}.}
\label{fig:10pmap}
\end{figure}

As the masses of the SUSY particles increase, acceptable relic density
is obtained by enhancing the annihilation rate. This happens in a
narrow strip in the $m_0$-$m_{1/2}$ plane (the co-annihilation tail),
where the $\tilde{\chi}^0_1 \tilde{\tau} \to \tau \gamma$ co-annihilation process
becomes enhanced due to near degeneracy between the masses of the
$\tilde{\tau}$ and $\tilde{\chi}^0_1$ LSP. The precision of the WMAP
measurement of $\Omega_{\chi} h^2$ has reduced the cosmologically
allowed region to a very narrow one for each value of $\tan \beta$
and sign of $\mu$. The upper tip of this region, corresponding to the
heaviest supersymmetric particle spectrum, is defined by its intercept
with the slepton LSP ($m_{\tilde{\tau}} < m_{\chi}$) boundary. This is
typically located at $m_{1/2} \sim \cal{O}$(1~TeV). As we shall see
below, this region can be entirely covered by LHC searches.

\subsubsection{Focus Point Region}

\label{sec:cmssm3}

In the focus point region at high $m_0$ the LSP changes its
nature. Instead of being dominantly gaugino, as is the case elsewhere,
it acquires a significant Higgsino content as $\mu$ is driven to small
values \cite{Feng:1999mn}.  At high values of $\tgbet$ the region
where this occurs can lie at rather low values of $\mhlf$. As the
Higgsino component can have a coupling to the SM gauge bosons, which
is forbidden for the bino component by gauge invariance, $\sigma$ is
enhanced \cite{Feng:2000zu}. Furthermore the $\chioi$ becomes nearly
mass degenerate with the $\chipm$ and $\chioii$. Consequently a number
of additional annihilation and co-annihilation processes contribute
which together can reduce the $\chioi$ relic density further to levels
compatible with CMB data. Unlike the co-annihilation tail, the focus
point region can extend to very large values of $m_0$ leading to
squarks, sleptons and gluinos that are too heavy to be observed at
LHC.

The very existence of a cosmologically acceptable focus point region
in CMSSM parameter space is sensitively dependent on the parameters
describing the model, especially the top quark mass. In addition the
large value of $m_0$ results in a ``fine-tuning'' problem; a delicate
cancellation is required between different contributions to the Higgs
potential to ensure that $M_Z<< m_0$. This problem worsens as $m_0$
increases.

\begin{figure}
\begin{center}
\epsfig{file=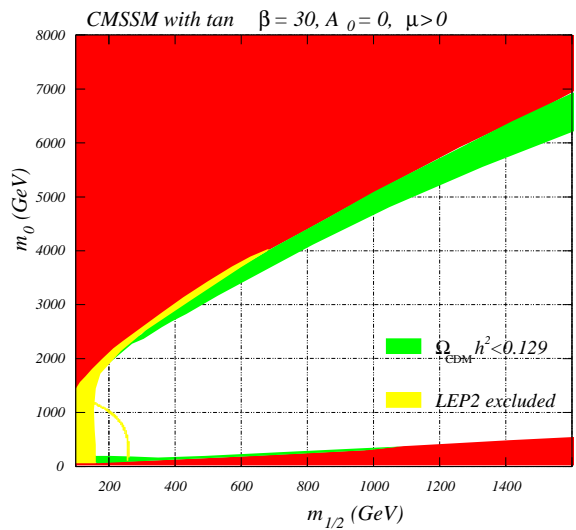,width=7.5cm,height=7.0cm,clip} 
\end{center}
\caption{The focus point region in the CMSSM $m_0-m_{1/2}$ plane for
$A_0=0, \tan\beta =10, \mu>0$. The medium (green) region is consistent
with WMAP relic density constraints ($0.094 < \Omega_{\chi} h^2 <
0.129$), while the light (yellow) region is excluded by LEP2
bounds. From Ref.~\cite{Baer:2003wx}. See also
Ref.~~\cite{Chattopadhyay:2003xi}.
\label{focus}}
\end{figure}

\subsubsection{Higgs Pole Region}

\label{sec:cmssm2}

When the value of $\tan \beta$ exceeds $\simeq 30$ another opportunity
for enhanced $\chioi$ annihilation is offered by the presence of heavy
neutral Higgs particles, $A^0$ and $H^0$ with masses $m_A$ such that
$m_{\chioi} \simeq m_{A} / 2$. In this case the annihilation of
lightest neutralinos is enhanced through resonant (s-channel) heavy
Higgs exchange \cite{Drees:1992am}.  These processes can allow CMSSM
models with $\mo$ and/or $\mhlf$ of order 1 TeV or greater to satisfy
relic density constraints. The cosmologically acceptable region in
these models takes the form of a ``funnel'' pointing toward large
$m_A$ defined by relic density contours passing around the
annihilation pole. For values of  $\tan \beta \gtap 60$ the coupling
of the Higgs boson to bottom quarks becomes large and reliable
computations are not possible.

\subsection{Beyond the CMSSM}

\label{sec:beyond}

\begin{figure}
\epsfig{file=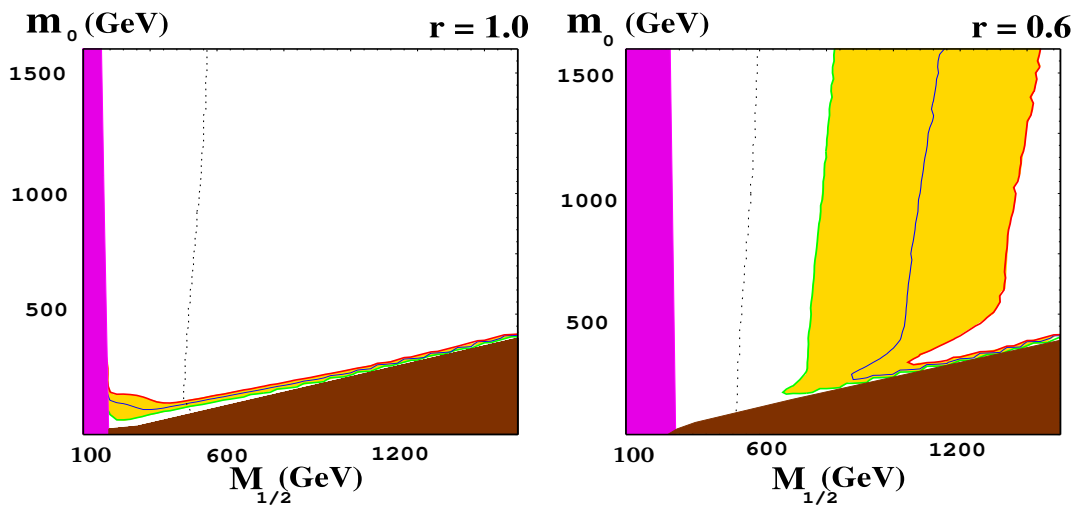,width=14.5cm,height=7.0cm} 
\caption{ $\chioi$ LSP relic density in the $m_0-m_{1/2}$ plane for
$\tan\beta = 5$ and differing gaugino mass assumptions. The
cosmologically allowed relic density regions (light (yellow and green)
shading) are displayed for $r=1$ (left panel) and for $r=0.6$ (right
panel).  Dark (brown) shaded regions in the lower right are excluded
due to a charged LSP.  Medium (pink) shaded regions on the left are
excluded by LEP bounds on the chargino mass. The Higgs mass contour of
$m_h = 113$~GeV is given by the dotted vertical line. Adapted from
Ref.~\cite{Birkedal-Hansen:2002am})}
\label{hansen}
\end{figure}

As the assumptions that lie behind the SUSY breaking in the CMSSM are
relaxed, the range of SUSY particle masses for which there exists an
acceptable dark matter candidate is considerably expanded.  Within the
CMSSM scheme, the assumption of the unification of the gluino and
gaugino masses at a high scale ensures that the neutralino LSP is
predominantly bino in the bulk region. If the LSP were dominantly wino
on the other hand, it would have larger couplings and larger
annihilation rates. Its mass could therefore be larger. This is
illustrated in Figure~\ref{hansen} from
Ref.~\cite{Birkedal-Hansen:2002am} which shows the effect of changing
$r=M_2/M_1$ from its unification value of $r=1$. For $r=0.6$ the LSP
becomes wino-like and the bulk region expands as shown in the right
hand plot.

The CMSSM also assumes that the soft scalar masses have a common value
at the unification scale. This assumption can also be relaxed leading
to modified relations between the squark, slepton and Higgs boson
masses. This causes the LEP Higgs mass limit to no longer directly
constrain the slepton and squark masses. The bulk region is not
greatly affected by this modification, and the co-annihilation tail
still exists, however the position and character of the focus point
region can be radically altered since the squark and slepton masses no
longer have to be large to obtain a small value of $\mu$
\cite{Baer:2003jb}.

Another mechanism of SUSY breaking, ``anomaly mediation'' (AMSB)
\cite{Randall:1998uk}, results in a SUSY spectrum where the LSP is a
wino. The simplest version of this class of models is ruled out as it
predicts tachyonic sleptons, however variants can escape this
constraint \cite{Harnik:2002et}. The characteristic feature of AMSB
models is the near degeneracy of the LSP and the charged wino
resulting in a large annihilation cross-section. Consequently if the
LSP has a mass around 100~GeV far too few survive today to make a
significant contribution to the dark matter. Nevertheless if the mass
of the LSP is very large it is possible that a satisfactory dark
matter density could be achieved.

In addition to gravity mediated models such as the CMSSM there also
exists a class of gauge mediated SUSY breaking (GMSB)
models\cite{GMSB, GMSBrev} in which the supersymmetry breaking is
mediated by gauge interactions. GMSB models assume that supersymmetry
is broken with a scale $\sqrt{F}$ in a sector of the theory which
contains heavy non-SM particles. This sector then couples to a set of
particles with SM interactions, called messengers, which have a mass
of order $M$. The mass splitting between the superpartners in the
messenger multiplets is controlled by $\sqrt{F}$. One (two) loop
graphs involving these messenger fields then give mass to the
superpartners of the gauge bosons (quarks and leptons) of the SM. GMSB
models are preferred by some because the superpartners of the Standard
Model particles get their masses via gauge interactions, so there are
no flavor changing neutral currents, which can be problematic in the
CMSSM. The non-observation of supersymmetric particles and new
particles associated with the messenger sector provides a bound on the
SUSY breaking scale: $F\gtap \frac{4\pi}{\alpha_2}m_{\tchi_2^+} M$
GeV$^2$ where $\alpha_2\sim 1/28$ is the coupling strength of the
$SU(2)$ electroweak interactions. Using $M,m_{\tchi_2^+}> 100$~GeV
implies $\sqrt{F} \gtap 2$ TeV however in most models $F$ is at least
ten times larger than this. Note that for gauge mediation to dominate
$F$ cannot be too large.

In GMSB models the characteristic spectra of superparticles are
different from those found in CMSSM models. In particular the lightest
supersymmetric particle is now the gravitino ($\tilde{G}$) whose mass
is given by $m_{\tilde{G}}\sim \frac{\sqrt{F}}{100 \hbox{TeV}}$ eV,
implying that $m_{\tilde{G}}> 0.2$eV. The gravitino can have a mass of
up to a GeV or so in specific models. It has feeble couplings
and can be produced with significant rates only in the decays of
particles which have no other decay channels. Its interactions are so
weak that annihilation in the early universe typically only occurs via
gravitational interactions and it falls very rapidly out of thermal
equilibrium \cite{Moroi:1993mb} leading to overclosure of the
universe. This problem can be avoided if the temperature to which the
universe is reheated after the end of inflation is sufficiently low to
avoid over-production \cite{Bolz:2000fu}. An alternative source of
gravitino dark matter is non-thermal production in which the next to
lightest SUSY particle (NLSP), which is usually a slepton or bino,
decays to the gravitino. The lifetime of the NLSP can be sufficiently
long for it to be effectively stable as the universe evolves, decaying
only later to produce gravitino dark matter, long after it has fallen
out of thermal equilibrium \cite{Ellis:2003dn}.

In certain SUSY models it is possible for the LSP to be a
sneutrino. This is however not favored in most models as the
sneutrinos are usually heavier than the right handed sleptons due to
the their larger gauge couplings. Nevertheless a stable sneutrino is a
good dark matter candidate\cite{Ibanez:kw} providing its mass is
around 500~GeV~\cite{Falk:1994es}.  Interaction rates of sneutrinos in
direct detection experiments (see Section~\ref{sec:direct}) are
expected to be similar to those of heavy neutrinos of the same
mass. Consequently current bounds from these experiments rule out
sneutrino dark matter for sneutrino masses less than 1 TeV
\cite{Mori:1992yq}.

There are a small number of SUSY models in which the gluino is the
LSP \cite{Raby:1998xr}, however these are rare since ordinarily the larger coupling
strength of QCD drives the gluino to larger masses than the other
gauginos. Gluino LSPs produced thermally in the early universe are
also unlikely to contribute significantly to the dark matter because
they annihilate very efficiently via strong interactions leading to a
number density typically of order $10^{-10}$ that of baryons.  Gluinos
cannot therefore account for a significant part of the dark matter
unless they are produced non-thermally.

So far we have mostly assumed that the LSP was initially in thermal
equilibrium. If this was not the case, then the prediction of its
relic abundance is no longer valid. In particular it has been pointed
out that the decay of ``moduli fields'' expected in some unification
models can produce too many LSPs \cite{Coughlan:1983ci}.  In gauge
mediated scenarios, the excess production of gravitinos restricts the
range of allowed masses\cite{Asaka:1997rv}. In the case of anomaly
mediation, it has been suggested that this mechanism can produce
sufficient LSPs to obtain the correct relic density for LSP masses
around 100~GeV~\cite{Moroi:1999zb}.

\section{LHC measurements of SUSY Dark Matter Properties}

\label{sec:lhc}

\subsection{Inclusive Searches}

\label{sec:lhc_incl}

\begin{figure}
\begin{center}
\begin{tabular}{c c}
\epsfig{file=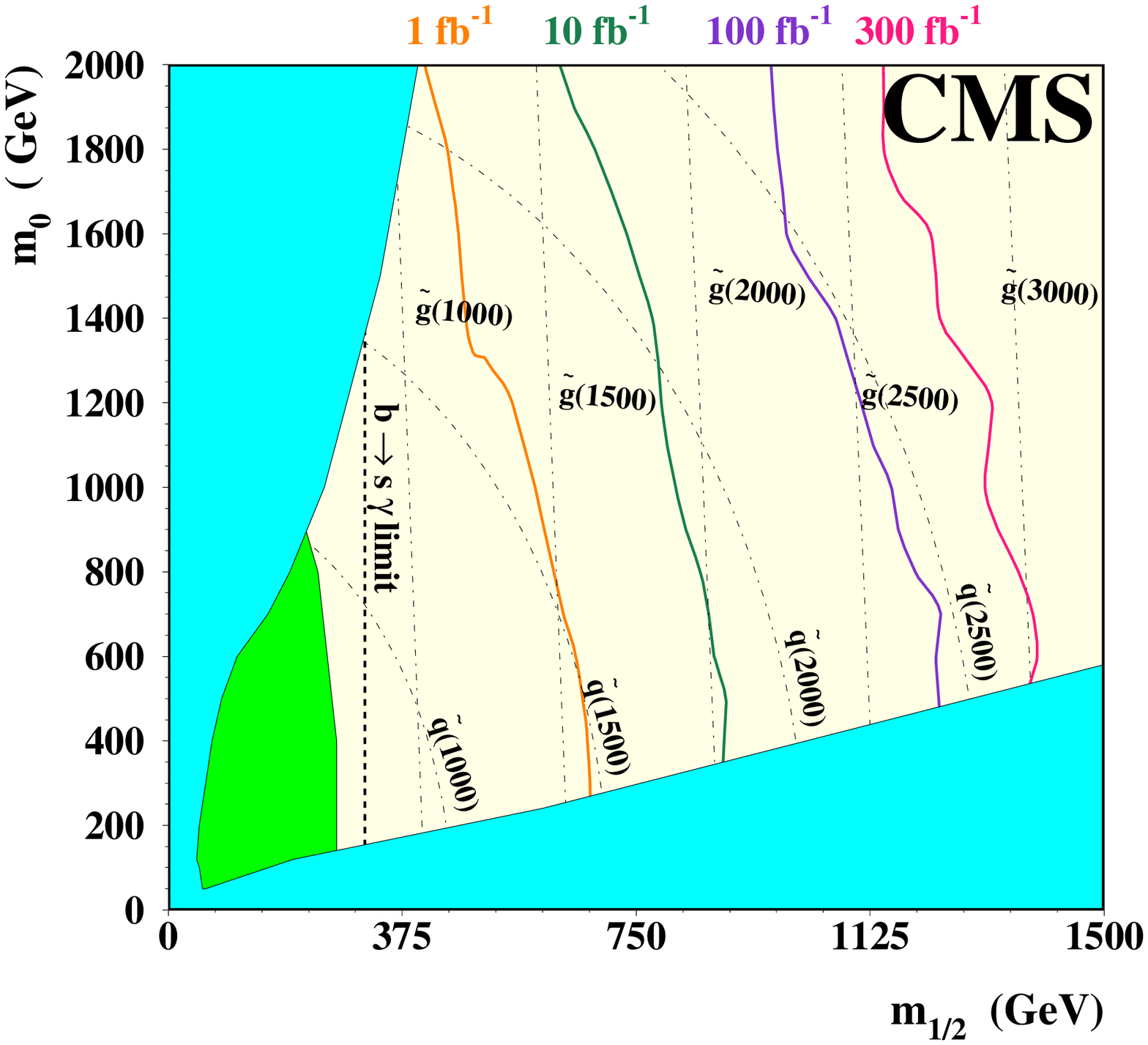,width=8.0cm} &
\epsfig{file=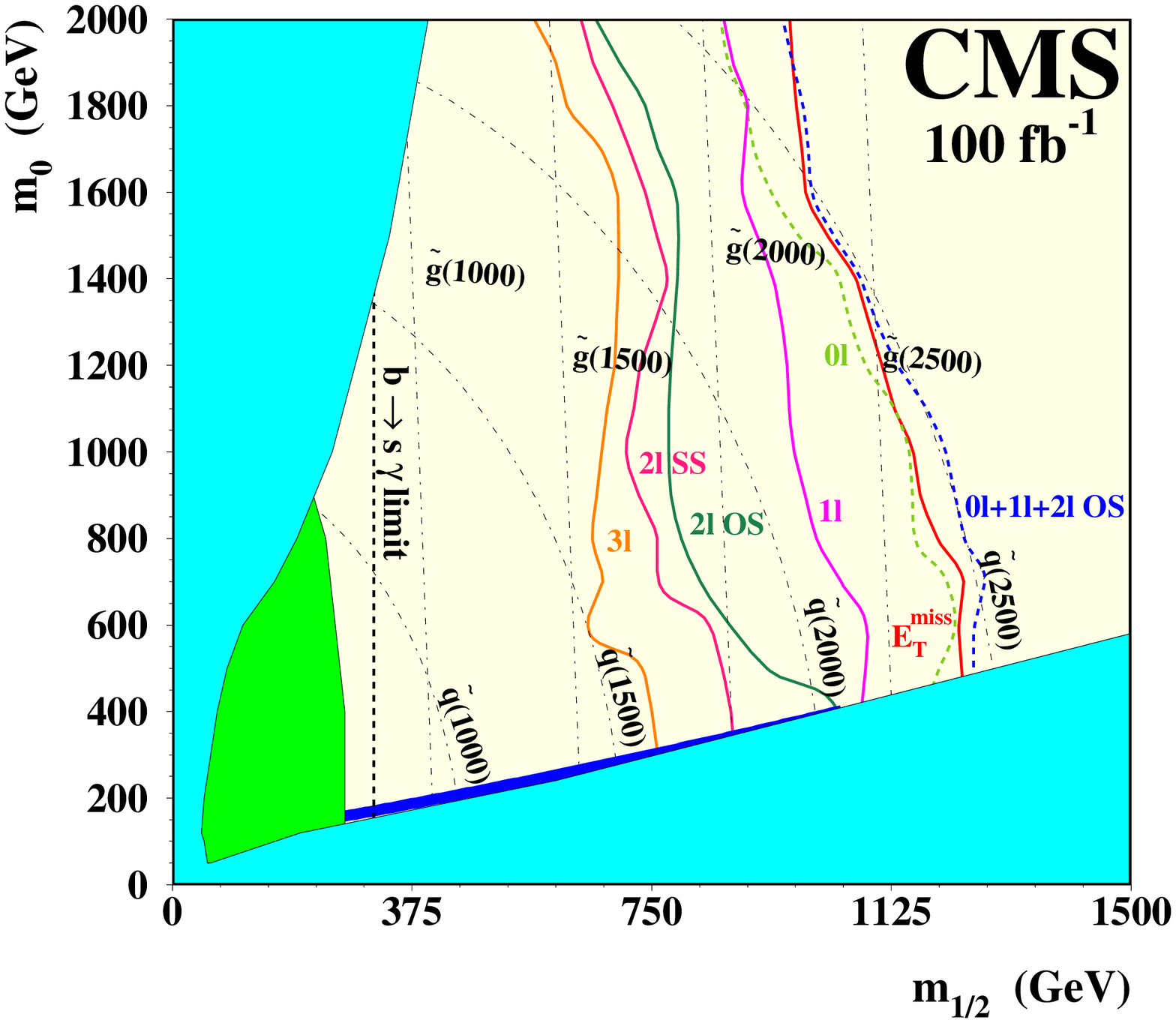,width=8.0cm} \\
\end{tabular}
\end{center}
\caption{Sensitivity of LHC experiments to supersymmetry in the CMSSM
$m_0$-$m_{1/2}$ plane. Left: reach of the $Jets~+~E_T^{miss}$ search
channel to obtain a 5~$\sigma$ sensitivity for different integrated
luminosities shown as solid lines. The dashed lines show fixed values
of the squark and gluino masses. Right: reach of different search channels for
100~fb$^{-1}$ of data indicated by the solid lines The cosmologically preferred region is
indicated by the dark (blue) line extending over the excluded region
where the LSP is charged (adapted from Ref.~\cite{Abdullin:1998pm}).}
\label{fig:scan}
\end{figure}

The first issue is the reach of the LHC experiments for
discovering supersymmetry.  The reach is dependent on the details of
the SUSY model, however the characteristic signals of missing
transverse energy (from invisible LSPs), a large multiplicity of
hadronic jets (from decays of squarks and gluinos) and/or isolated
leptons (from decays of sleptons and gauginos) are sufficient to
ensure detection over a large mass range. The left plot in
Figure~\ref{fig:scan} shows the expected sensitivity of the CMS
detector via the $Jets$~+~$E_T^{miss}$ search channel to models in the
CMSSM $m_0$-$m_{1/2}$ plane as a function of the integrated
luminosity.  About 30~fb$^{-1}$ of data ensure that a signal can be
observed over the whole extension of the co-annihilation tail. The
right plot of Figure~\ref{fig:scan} shows the sensitivity with
100~fb$^{-1}$, in a variety of channels characterised by the number of
observed isolated leptons. The most sensitive channel is that with
jets and missing transverse energy and no isolated leptons labelled
``0l''. Signals with two isolated leptons of the same charge (marked
``2l SS'') can arise from the pair production of gluinos and their
subsequent decay to charginos. Since the gluino is a Majorana fermion,
it can decay into charginos of either charge. Their decays then give
rise to this signal.  Signals with two isolated leptons of opposite
charge (marked ``2l OS'') arise from the same source but also from
decays of the type $\tchi_2^0 \to \ell \ell \lsp$. Trilepton events
(marked ``3l'') arise predominantly from events containing $\tchi_2^0$
and a chargino.

\subsection{Measurement of Cold Dark Matter Properties}

\label{sec:lhc_excl}

The observation of signals characterized by large missing transverse
energy, jets and/or isolated leptons at the LHC will give a strong
indication that supersymmetry or some other New Physics resembling
SUSY such as UED (Section~\ref{sec:ued}) is present. It will also
provide circumstantial evidence that $R$ parity is conserved. The
limited sensitivity of LHC experiments to the LSP lifetime
($\tau_{\chi}$ $\gtrsim$ 1 ms \cite{atltdr-susy}) means that final
confirmation that cold dark matter is supersymmetric in nature will
require observation of signals in astroparticle experiments consistent
with mass and cross-section predictions derived from LHC
measurements. This is especially important given that the dark matter
relic density measured by CMB experiments could consist of
non-supersymmetric components in addition to that provided by the
LSP. It will therefore be vital to measure sufficient SUSY parameters
to verify consistency of the observed signals not only with CMB data
but also with signals in these non-accelerator experiments.

Initially these measurements will be performed within the context of a
specific SUSY breaking model, such as the CMSSM, by comparing
significances in the various inclusive
$Jets$~+~$E_T^{miss}$~+~$n$~$leptons$ channels discussed above. As
more data accumulates, it will be possible to begin to reconstruct
SUSY particle decays. This may enable model-independent measurement of
the masses of some SUSY particles. However this will not necessarily
be sufficient to provide a model-independent prediction of
the dark matter relic density or other dark matter properties. This is
because we do not know {\it a priori} which annihilation and
co-annihilation processes dominated in the early universe, or
equivalently in which region of which parameter space Nature's chosen
model lies. With only limited knowledge of the SUSY mass spectrum it
will therefore be necessary initially to assume a specific SUSY
breaking model. This will enable the full SUSY mass spectrum to be
obtained from a limited set of measurements, with additional input
coming from the unification assumptions. Only once detailed
model-independent measurements of more of the individual SUSY masses
and couplings have been made can this model-dependency in the
estimates of dark matter properties be reduced or eliminated.

As SUSY events will each contain two invisible LSP's, it will not
possible to fully reconstruct the final states in these
events. Measurements using exclusive channels will therefore focus on
identifying key features of the SUSY particle decays. This typically
involves measuring the end-points and shapes of invariant mass
distributions of leptons and jets arising in the decay of SUSY states
\cite{atltdr-susy,fep97a,hb00a}. Measurement of CMSSM model parameters
can be accomplished through a global fit of the parameters to the
positions of these edges, while model-independent mass measurements
can be obtained by solving the various mass relations simultaneously.

The exclusive measurements which can be used to estimate dark matter
properties in either a model-dependent or model-independent fashion
depend strongly on the (co-)annihilation processes which drove the
reduction of the dark matter density in the early universe. Within the
context of the CMSSM the regions of parameter space where these
different processes dominate correspond those discussed in
Section~\ref{sec:cmssm}. We shall therefore now discuss the
measurements which can be made in each of these regions.

\subsubsection{Bulk Region and Co-Annihilation Tail}

\label{sec:lhc_excl1}

\begin{figure}
\begin{center}
\begin{tabular}{c c}
\epsfig{file=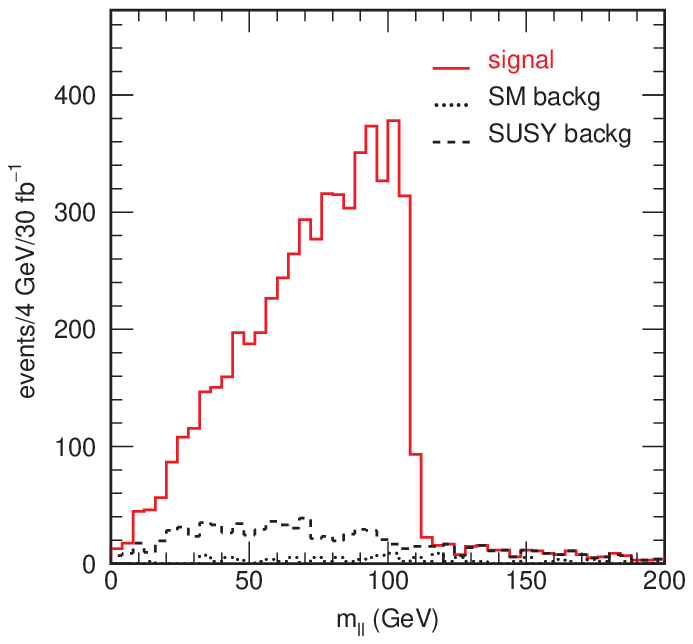,width=7.5cm,height=7.0cm} &
\epsfig{file=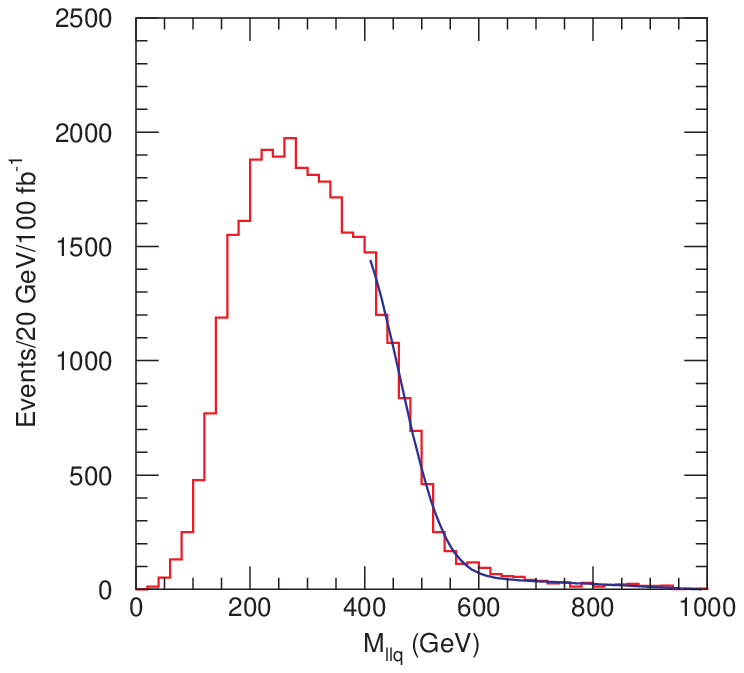,width=7.5cm,height=7.0cm} \\
\end{tabular}
\end{center}
\caption{ Invariant mass distributions obtained from the LHCC Point 5
 benchmark model ($m_0 = 100$~GeV, $m_{1/2}=300$~GeV, $A_0=300$~GeV,
 $\tgbet=2.1$, $\mu>0$) for 100~fb$^{-1}$ of data. Left: the invariant
 mass of a pair of opposite sign same flavor leptons arising from
 squark decay. Right: the invariant mass of a pair of opposite sign
 same flavor leptons and jet arising from squark decay. From
 Ref.~\cite{atltdr-susy}.}
\label{susy-edge}
\end{figure}

In the bulk and co-annihilation regions of CMSSM parameter space the
mass of at least one of the sleptons or staus lies between that of the
$\chioii$ and the $\chioi$ (LSP). The production of supersymmetric
particles is dominated by squarks and gluinos as these have strong
interaction couplings. If gluinos are heavier than squarks then they
decay to squarks. In these regions there is a significant decay rate
for $\tilde{q}\to q \tilde{\chi}_2^0 \to q \tilde{\ell}^{\pm} \ell^\mp
\to q \ell^\pm \ell^\mp \lsp$ and/or the equivalent stau channel. This
decay chain gives rise to a jet, from the quark, a dilepton pair of
the same flavor and opposite charge, and missing energy, from the
$\lsp$. After events are selected to contain a pair of isolated
leptons, jets and missing energy, the left plot in
Figure~\ref{susy-edge} shows the dilepton invariant mass distribution
for a typical model. A clean structure is visible with a kinematic
end-point arising from this decay chain. Events beyond the end-point
are due to other SUSY particle decays. The position of this end-point
is determined by the masses of $\tilde{\ell}^{\pm}$,
$\tilde{\chi}_2^0$ and $\lsp$. If events are selected with a dilepton
pair below this end-point and this pair is then combined with all
possible jets in the event, the mass distribution of the jet-dilepton
combination with the lowest mass displays a kinematic structure whose
end-point depends also on the quark mass (Figure~\ref{susy-edge}
(right)). There is also a minimum value of the mass of the dilepton
and jet and an upper value of the mass of a jet and single lepton.

\begin{figure}
\begin{center}
\epsfig{file=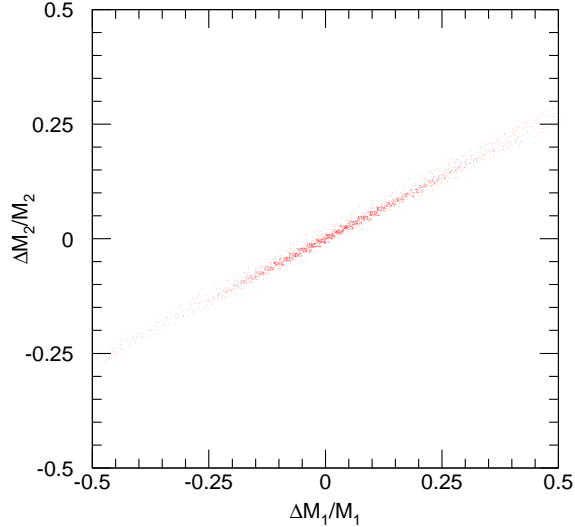,width=7.5cm,height=7.0cm} 
\end{center}
\caption{Plot showing errors on masses determined from the
distributions in Figure~\ref{susy-edge}. Note the strong correlation.
From Ref.~\cite{atltdr-susy}.}
\label{mass-correl}
\end{figure}

These four quantities can be used to solve for the four masses:
$\tilde{q}$, $\tilde{\ell}^{\pm}$, $\tilde{\chi}_2^0$ and $\lsp$. It
is important to note that a model is not required; simply the decay
chain must be identified. When the masses are extracted precisions
ranging from $\sim$ 3 \% for the $\sql$ to $\lesssim$ 14 \% for the
$\chioi$ for the LHCC Point 5 benchmark model are obtained
\cite{Allanach:2000kt}. Note however that the errors are strongly
correlated as is illustrated in Figure~\ref{mass-correl}: an
independent measurement of one mass improves the precision on all of
the others. Note also that when stau co-annihilation dominates, the
above signatures can involve very soft taus which can be difficult to
observe or measure with significant accuracy. Nevertheless if the mass
difference between the $\staui$ and either of the lighter neutralinos
is more than a few GeV then it should be possible to make at least a
rough estimate of the relevant end-point positions.

\begin{figure}
\begin{center}
\epsfig{file=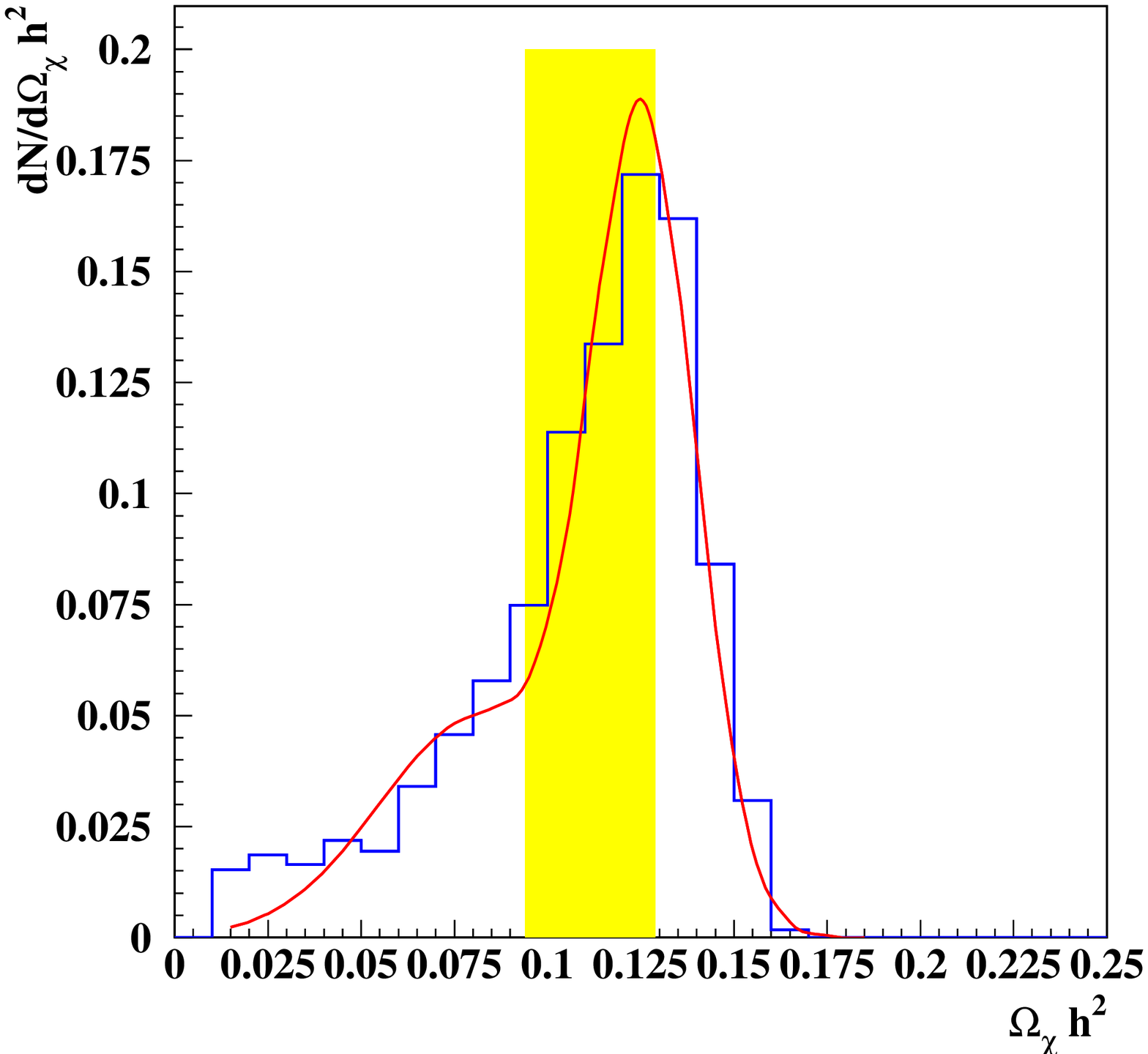,width=8.0cm}\epsfig{file=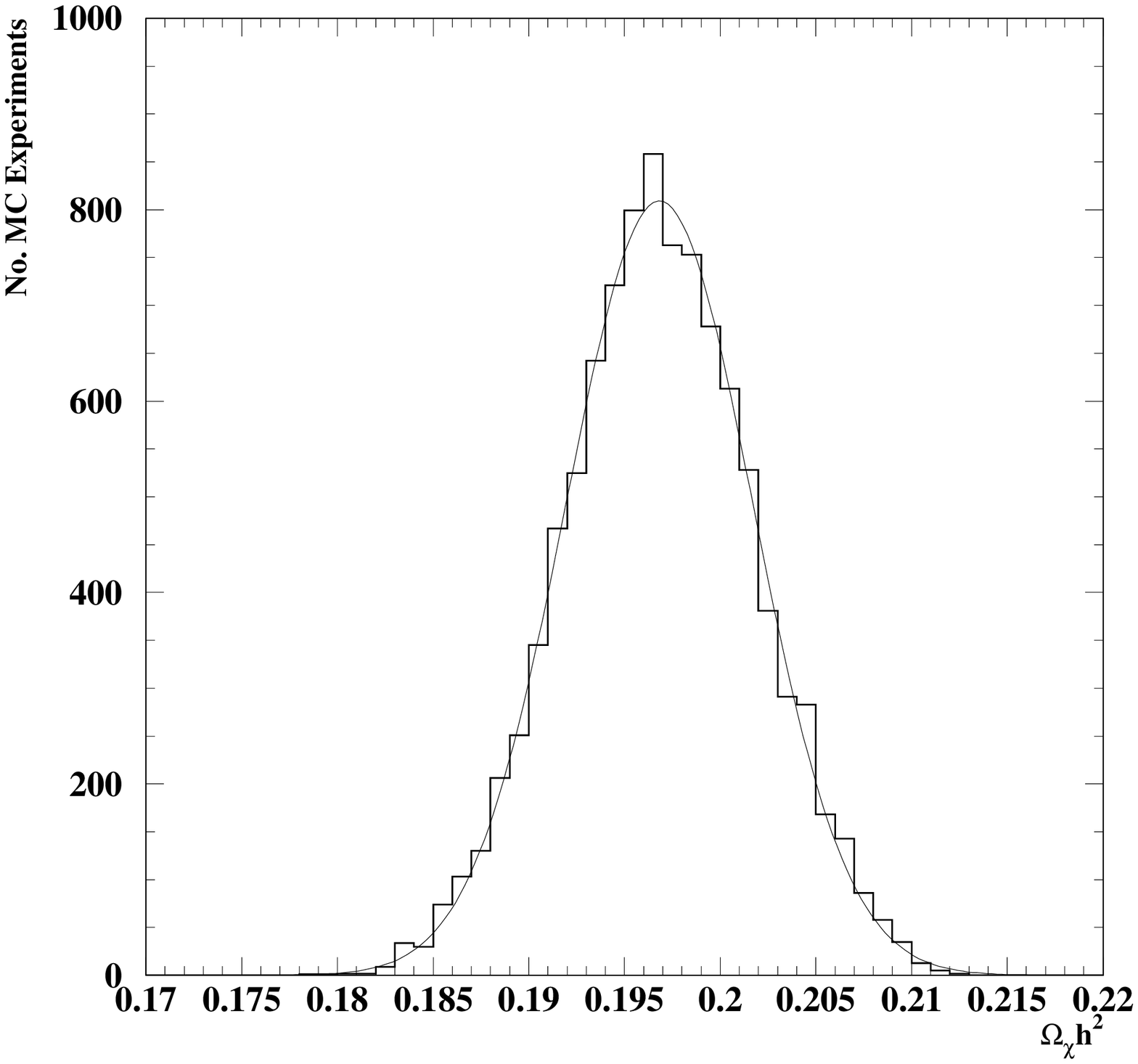,width=8.0cm}
\end{center}
\caption{Probability density functions for the dark matter density as
can be determined at the LHC. The left-hand distribution shows dark
matter densities calculated from post-LEP benchmark model B'
\cite{Battaglia:2003ab} by first fitting to invariant mass edges with
model-independent SUSY particle masses, and then fitting CMSSM
parameters to the measured mass distributions, before using those
parameter values to calculate $\ohsq$. The range $0.094 < \Omega_\chi
h^2 < 0.129$, favoured by WMAP, is shown for comparison. The
right-hand distribution was obtained from the SPS1a benchmark
model~\cite{Allanach:2002nj} by directly fitting CMSSM parameters to
the invariant mass edges before using those parameter values to
calculate $\ohsq$ (from Ref.~\cite{Polesello:2004qy}).}
\label{fig:fit}
\end{figure}

Initial model-dependent (see Section~\ref{sec:lhc_excl}) predictions
for dark matter properties can be obtained from measurements such as
these in two different ways. One approach is to fit the parameters of
the assumed SUSY breaking model to the measured model-independent SUSY
particle masses. The model parameters thus obtained can then be varied
within their statistical and systematic errors to produce a
probability density function for {\it e.g.} $\ohsq$ (left hand plot in
Figure~\ref{fig:fit}). This approach however does not take into
account the correlations between the errors on the different
measurements. An alternative approach which does take these
correlations into account is to perform a global fit of the model
parameters directly to all the end-point measurements prior to
calculating the $\ohsq$ PDF (right hand plot in
Figure~\ref{fig:fit}). For the SPS1a bulk region benchmark model
\cite{Allanach:2002nj} a recent study \cite{Polesello:2004qy} has
shown that with 300 fb$^{-1}$ of data statistical measurement
precisions $\sim$ 2\%, 0.6\%, 9\% and 16\% can be obtained for the
CMSSM parameters $m_0$, $m_{1/2}$, $\tgbet$ and $A_0$
respectively. The sign of $\mu$ is also well-determined from the
fit. It is estimated that $\ohsq$ can be pedicted with a
statistical(systematic) precision $\sim$ 2.8\% (3.0\%). Further dark
matter properties such as the spin-independent $\chioi$-nucleon
elastic scattering cross-section $\sigma_{si}$ (see
Section~\ref{sec:direct}) are estimated with statistical precisions
$\lesssim$ 1\%, far smaller than the estimated factor $\gtrsim$ 2
systematic uncertainties in these quantities.

Full model-independent estimation of the dark matter density will
require use of the model-independent mass values discussed above
together with other measurements of quantities such as the masses of
the lighter staus, heavy Higgs bosons and heavier neutralinos in order
to show that slepton or stau (co-)annihilation dominates in the early
universe (i.e. that the model does not lie near the heavy Higgs pole,
or in the Focus Point region). More work on this subject is 
required.

\subsubsection{Focus Point Region}

\label{sec:lhc_excl2}

The phenomenology of the focus point region differs from that of the
bulk and co-annihilation regions primarily due to the large masses of the 
sfermions. Depending on the parameters, the LHC may perform 
measurements providing some constraints on the dark matter density, 
or only observe an excess of events due to supersymmetry, 
or even, at the upper end of the parameter range, not observe 
supersymmetry at all. Sequential two-body, slepton-mediated, decays 
of $\chioii$ to $\chioi$, which form the basis of the measurements 
discussed previously, are kinematically suppressed. Some other 
signatures may still observed in these models, depending on the value 
of $\Delta m = m_{\chioii}-m_{\chioi}$.

In models with $\Delta m < m_Z$, such as the LHCC Point 4 model ($\mo$
= 800~GeV, $\mhlf$ = 200~GeV, $\tgbet$ = 10, $\mu>0$, $\ao$ = 0; not
strictly speaking a focus point model but a good example of a model
with $\mo>>\mhlf$) studied in Ref.~\cite{fg-97,atltdr-susy}, the
$\chioii$ decays almost exclusively to three-body final states
involving a $\chioi$ and two opposite-sign same-flavor leptons or
quarks. The dilepton signature is easiest to identify and provides an
invariant mass end-point sensitive to $\Delta m$ which can be measured
with flavour subtraction techniques~\cite{atltdr-susy}. Combination of
the dilepton pair with a hard jet may also give sensitivity to the
$\sql$ mass while $\mo$ is sufficiently small for $\sql$ production to
be significant.  The mixed Higgsino-gaugino nature of the heavier
neutralinos and charginos leads to significant couplings to the $Z^0$,
and $h^0$ bosons, giving a characteristic $Z^0$ peak in the dilepton
mass spectrum. The production cross-section and $p_T$ distribution of
these $Z^0$ bosons can be used to constrain heavier neutralino and
chargino masses~\cite{atltdr-susy}. Further mass constraints were also
obtained from a dijet invariant mass end-point sensitive to
$m_{\glss}-m_{\chioii}$, albeit with potentially large combinatorics.
Sensitivity to squark and other sfermion masses is however difficult
to obtain.

Model-dependent techniques, such as the analysis of the shape of the
$E_T^{sum}$ distribution of SUSY events, are then required. A combined
fit of CMSSM parameters to the results of all the available
measurements for LHCC Point 4, gave a precision $\sim$ 6\% for $\mo$,
$\sim$ 1\% for $\mhlf$ and $\sim$ 15\% for $\tgbet$
(300~fb$^{-1}$). Unsurprisingly, the $\mo$ measurement precision was
relatively poor compared with that for $\mhlf$ due to the difficulty
of identifying and measuring the sfermion signals. The lack of
sensitivity to $A_0$ and sign($\mu$) in this study indicates that only
an order of magnitude estimate of $\ohsq$ could be obtained.

In the parameter region where $\Delta m > m_Z$ two-body decays of
$\chioii$ to $\chioi+Z^0$ and, for $\Delta m > m_h$, $\chioi+h$ become
accessible.  Phase-space suppression of the three-body $\chioii$ modes
and dominant $\chioii$ production in complex $\glss$ three-body decays
involving heavy flavors means that, even if an inclusive signal is
observed, it is difficult to find mass distributions providing
measurable end-points with little combinatorial background. Parameter
measurements must therefore rely upon other techniques involving the
$Z^0$ or $h$ signatures. This is an area where further study is still
required to assess the LHC potential.

The large masses of the strongly interacting particle partners cause
an increasing fraction of the total SUSY production cross-section to
be taken up by direct production of neutralinos and charginos.  These
events are characterised by little jet activity with instead
multi-leptons, or soft b-jets, arising for instance from $Z^0$ or $h$
decay. Interest in specific searches for these signatures has been
re-awakened by results of a recent study~\cite{Baer:2003ru}.  It
indicates that the reach in the focus point region of gaugino searches
at a 1~TeV $e^+e^-$ linear collider, where gaugino pair production can be
detected easily, can be greater than that of the LHC, when using the
conventional jets + $E_T^{miss}$ channels. At the LHC the production
cross-sections for the electroweak gauginos in the focus point region
are comparatively large, and the SM backgrounds, with appropriate
cuts, can be made much smaller.  The disadvantage of these channels is
that hard cuts on jet activity are required to reject hadronic
backgrounds leading to low efficiency and potentially large
systematics due to poor understanding of mini-jet backgrounds from
pile-up and the underlying event.
These searches resemble those performed at the Tevatron. The characteristic
signatures are dileptons or trileptons + $E_T^{miss}$ arising from the
production of electroweak gauge bosons in gaugino decay. In an early
study of these channels~\cite{Baer:1995va} a plausible central jet
$p_T$ cut of 25~GeV was assumed, consistent with that used in more
recent central jet veto studies for Higgs searches in the vector boson 
fusion channel~\cite{Asai:2004ws}. Only limited sensitivity to the high 
$m_0$ region was found. However with a more refined analysis and a finer 
granularity scan of the narrow focus point region it may be possible 
to improve the reach of the method. 

At very large values of $\mo$ and $\mhlf$, the $\chioi$ becomes almost
pure Higgsino and the $\chipm$ can be almost mass degenerate with it
(the so-called ``inversion region'' of the hyperbolic
branch~\cite{Chattopadhyay:2003xi}). In this scenario,
$m_{\chipm}-m_{\chioi}$ and $m_{\chioii}-m_{\chioi}$ are once again
small and the two-body gauge and Higgs boson decay channels are
closed.  Three-body decays again dominate, however the chargino and
neutralino mass differences can be such that only very low $p_T$
leptons and jets are produced (although the NLSP does decay inside the
detector~\cite{Chattopadhyay:2003xi}). Triggering upon such events is
likely to be very difficult at the LHC, and measurement of sparticle
masses still more so.
The limited scope for SUSY particle mass measurement in this region 
means that estimation of the $\chioi$ relic density is a difficult 
task. This problem is compounded by the fact that in this scenario 
the value of $\ohsq$ is strongly dependent on the top mass, which 
will only be known to a systematics-limited accuracy $\sim$~2~GeV 
at the LHC, after one year of low luminosity running~\cite{atltdr-top}.

Following any SUSY discovery at the LHC, the first indications that
the focus point scenario is realised would likely come from a
comparison of signal significance in different inclusive channels
distinguished by the required number of leptons. In particular the
signal significances in the jets + $\etm$ + $n\geq1$ lepton channels
are enhanced relative to that in the $n=0$ leptons channel in the
focus point region due to the abundant production of heavy flavors in
gluino decay and leptons in chargino and neutralino decays via gauge
bosons.  Further evidence could then be provided by exclusive
channels.  At low $\Delta m$ evidence for a di-lepton/di-tau end-point
can be observed with a lepton/tau $p_T$ asymmetry consistent with
three-body decays~\cite{Abdullin:1998pm}, while at high $\Delta m$ no
evidence for such an end-point would be observable.

Relic density estimation within the CMSSM framework at low $\Delta m$
would require use of the mass measurements and model parameter fitting
techniques discussed above. At high $\Delta m$, $\mhlf$, $\tgbet$ and
sign($\mu$) could potentially be estimated from measurements involving
direct chargino/neutralino production. But without an accurate
estimate of $\mo$, the magnitude of $\ohsq$ would be difficult to
assess.

Estimation of the $\chioi$ relic density in a more model-independent
manner in the focus point scenario is even more difficult. Quantities
which would need to be measured include the neutralino mass matrix
parameters $M_1$, $M_2$, $\tgbet$ and $\mu$, and $m_A$. While it may
be possible to measure some or all of the neutralino/chargino sector
parameters from direct production studies (high $\Delta m$), the $A$
may well be inaccessible to direct study (via e.g. $H/A \ra \tau\tau$)
due to its very high mass at high $\mo$. However, if $H/A \ra
\tau\tau$ were observed then this would enable a rather accurate
measurement of $\tgbet$ to be performed~\cite{atltdr-higgs}.

\subsubsection{Higgs Pole Region}

\label{sec:lhc_excl3}

The potentially very large masses of the strongly interacting
super-partners in these ``rapid-annihilation'' models means that, if
realized in nature, their discovery at the LHC is by no means
guaranteed (see, for example, the benchmark model ``M'' in
Ref.~\cite{Battaglia:2003ab}). Nevertheless, if their spectrum is
within reach of the LHC, it may be possible to measure many of the
model parameters through the use of invariant mass end-point
techniques. The rapid-annihilation funnel region in general occurs in
the high $\tgbet$ $\mo-\mhlf$ plane relatively close (in all but the
highest $\tgbet$ scenarios) to the stau co-annihilation tail discussed
in Section~\ref{sec:lhc_excl1}. Consequently for many
rapid-annihilation models the $\staui$ is light and
$m_{\staui}<m_{\chioii}$. Ditau signatures may therefore be used to
identify SUSY events containing $\chioii$ decays, and fits to
invariant mass thresholds and end-points involving one or more of
these taus used to constrain either CMSSM parameters or
model-independent SUSY particle masses. This analysis should be
somewhat easier than in the stau co-annihilation case (for the same
total SUSY cross-section) since here the $\staui-\chioi$ mass
difference is greater and hence the taus from the $\staui$ decays have
greater $p_T$.

In addition to ditau production further signatures can be used to
identify and measure SUSY events in rapid-annihilation models. These
models in general possess relatively large values of $\mo$ compared
with those of models in the stau co-annihilation tail, and this
enhances the Higgsino content of the lighter neutralinos and hence the
branching ratio of $\chioii \ra Z^0/h + \chioi$. This allows further
constraints to be placed on masses through identification of $Z^0$ and
$h^0$ production in high $\etm$ events \cite{Hinchliffe:2001bz} and then
measurement of invariant mass end-points in $Z^0q$ and $hq$ invariant
mass distributions. Additional constraints can be obtained by
searching for $\sqr \ra \chioi q$ decays producing dijets + $\etm$,
which give sensitivity to the $\sqr$ mass
\cite{Baer:2003wx,Hinchliffe:2001bz}. Given an estimate of the mass of
the $\chioi$ from e.g. the ditau analysis the measurement can be
performed either by analyzing the shape of the $p_T$ distribution of
each of the jets \cite{Hinchliffe:2001bz,atltdr-susy} or by fitting to
the end-point of the distribution of the dijet $m_{T2}$ variable
\cite{Allanach:2000kt,fullsim}.

A case study of CMSSM parameter fitting in a model with this
phenomenology is the analysis of LHCC Point 6 ($\mo$ = 200~GeV,
$\mhlf$ = 200~GeV, $A_0=0$, $\tgbet=45$, $\mu<0$) described in
Refs.~\cite{Hinchliffe:1999zc,atltdr-susy}. By using a combination of
measurements of $m_h$ and $m_{\sqr}$, constraints on $m_{\glss}$ and
$m_{\glss}-m_{\tilde{b}_1}$ obtained from $\glss$ decays to
$\tilde{b}_1 b$, and fits to the ditau end-point, measurement
precisions $\sim$ 12\%, 5\% and 4\% were obtained for $\mo$, $\mhlf$
and $\tgbet$ respectively. However, the analysis was relatively
insensitive to $A_0$ and the sign of $\mu$, which would limit the
predictivity of these data in estimating the relic density.

In models with very large $\tgbet$ it is possible for the
rapid-annihilation funnel to be sufficiently wide for the
cosmologically-allowed region on the high $\mo$ side of the base of
the funnel to lie in the region where $m_{\staui}>m_{\chioii}$. The
phenomenology of such models resembles more closely that of the focus
point models discussed in Section~\ref{sec:lhc_excl2}, with
characteristic $Z^0/h + \etm$ ($\Delta m > m_Z$ or $m_h$) or ditau +
$\etm$ ($\Delta m < m_Z$) signatures arising from $\chioii$ decay.

\begin{figure}
\begin{center}
\epsfig{file=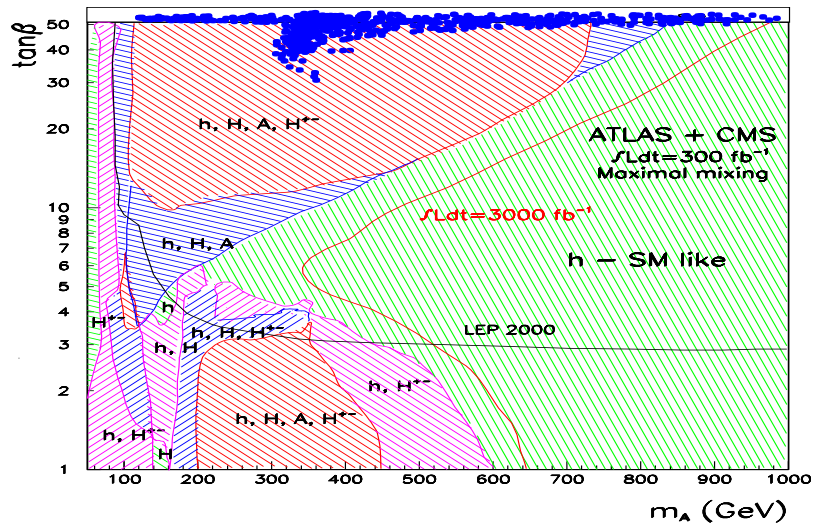,width=11.0cm,height=7.5cm,clip}
\end{center}
\caption[]{Estimated number of Higgs bosons observable at the LHC 
in the $M_A$, $\tan \beta$ plane. The cosmologically interesting solutions 
belonging to the $A$ funnel regions obtained with a parameter scan are 
shown by the dots. In the relevant region of parameters, the LHC will be 
essentially observing all the Higgs bosons 
(modified from~\cite{Gianotti:2002xx}).}  
\label{fig:matanb}
\end{figure}

Given a set of measurements of CMSSM parameters from the above
analyses it should be relatively straightforward to calculate the
$\chioi$ relic density, although the resulting error could well be
significant due to its strong dependence on both SUSY and SM
($m_{top}$) input parameters \cite{Ellis:2003cw}. In order to obtain a
more model-independent estimate of the relic density it will be
necessary first to confirm that $\chioi$ annihilation in the early
universe was dominated by resonant s-channel $A$ production. Assuming
that the $\chioi$ mass has been determined through the ditau + $\etm$
analysis $m_A$ must now be measured independently through a search for
e.g. $H/A \ra \tau\tau$ or (better) $H/A \ra \mu\mu$ to check that
$m_A \simeq 2m_{\chioi}$. It is interesting to observe that in the
CMSSM parameter region where a cosmologically interesting CDM density
is achieved due to the rapid $\chi \chi$ annihilation process through
the $A^0$ pole, the LHC should observe the $A^0$ and $H^0$ bosons for
masses at least as high as 800~GeV~\cite{cms-higgs}, corresponding to
$\mhlf \sim 950$~GeV with 300~fb$^{-1}$ integrated luminosity (see
Figure~\ref{fig:matanb}). An LHC upgrade, increasing the luminosity by
an order of magnitude, would ensure that the whole region is
covered. The $A$ mass measurement accuracy depends crucially on the
availability of the $A^0 \to \mu \mu$ channel in which case a
${\cal{O}}(0.1\%)$ precision can be reached with 300 fb$^{-1}$
($\delta m/m \sim$0.1\% for $m_A$ = 300~GeV at $\tan \beta$
=30)~\cite{atltdr-higgs}. In these cases $\tgbet$ could also be
measured to a precision $\sim$~5\%~\cite{atltdr-higgs}.  At the
largest values of $m_A$, where only the $A^0 \to \tau \tau$ is
available, the accuracy on $m_A$ will be typically of the order of a
few~\%, limited by energy calibration systematics.

\subsubsection{Other Scenarios}

\label{sec:lhc_excl4}

Measurements of SUSY parameters and hence dark matter properties in
other non-CMSSM scenarios have been studied in far less detail. If the
LSP is a wino then evidence could be provided by studying the
branching ratios of the $\chioii$ and $\chioi$, as well as measuring
the full neutralino mass matrix. Given a plausible model such as that
studied in Ref.~\cite{Birkedal-Hansen:2002am} it may then be possible
to measure sufficient parameters to estimate $\ohsq$. Similarly if the
LSP is a sneutrino then this will significantly alter the cascade
decay chains used in the CMSSM analysis but not prevent observation
of the kinematic end-points used to constrain masses or model
parameters. If a GMSB model is realised in nature then the prospects
for direct detection (Section~\ref{sec:direct}) are slim, but studies
at the LHC will be fruitful with a wide variety of signatures and mass
constraints accessible \cite{atltdr-susy}. This should enable the GMSB
model parameters to be estimated and, if the lifetime of
the NLSP can be measured, it should be possible to estimate the SUSY
breaking scale and hence the mass of the gravitino LSP.

\section{Cold Dark Matter and Extra Dimensions}

\label{sec:ued}

Extra dimensions offer an alternative solution to the
hierarchy problem. Instead of introducing a cancellation of the
divergences as in supersymmetry, extra dimensions bring the effective
unification scale from the Planck to the TeV scale. Of the several
models which have been proposed, Universal Extra Dimensions
\cite{Appelquist:2000nn} are of
particular interest here since they offer a cold dark matter
candidate. The simplest realisation of UED has all the SM particles
propagating in a single extra dimension of size $R$, which is
compactified on a $S_1/Z_2$ orbifold. This single parameter determines
the phenomenology of the model. Each Standard Model particle has an
infinite tower of Kaluza Klein (KK) excitations, with each level one
excitation possessing a mass equal to $1/R$ at tree level. This mass
degeneracy is broken only by radiative corrections. A further specific
feature of UED is the conservation of the Kaluza-Klein number at tree
level. Radiative
effects~\cite{Georgi:2000ks,vonGersdorff:2002as,Cheng:2002iz} break KK
number conservation down to a discrete conserved quantity known as KK
parity, given by $(-1)^n$ where $n$ is the KK level. The level one KK
states in UED models with conserved KK parity therefore strongly
resemble the particles appearing in SUSY models with conserved
R-parity and indeed it can be difficult to distinguish such models
experimentally. The lightest KK partner (LKP) at level one is often
neutral and has negative KK parity causing it to be stable on
cosmological time scales. The LKP is therefore a plausible dark matter
candidate again resembling LSP candidates.

\begin{figure}
\begin{center}
\epsfig{file=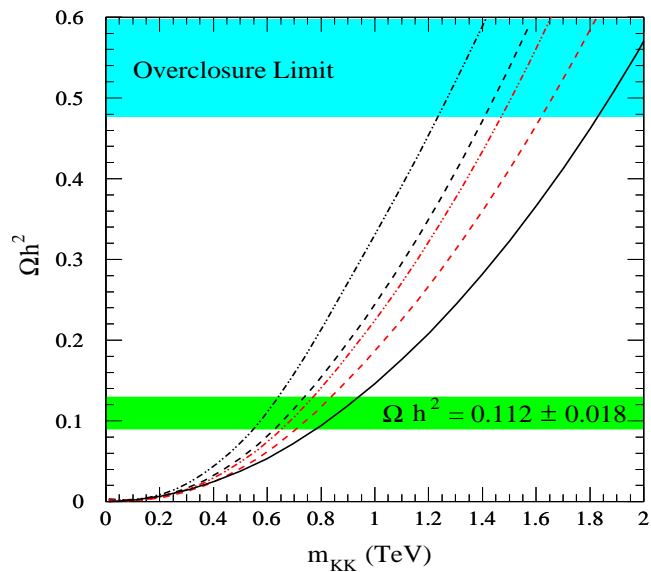,width=8.5cm,height=7.5cm}
\end{center}
\caption{Relic dark matter density as a function of the UED mass scale
$m_{KK}=1/R$ corresponding to the mass of the level one KK states. The
WMAP result is shown by the horizontal band.  The solid line includes
only the $B^{(1)}$ contribution. The dashed lines also include the
effect of co-annihilations with one and three generations of lepton
excitations for mass splittings of 1~\% and 5~\% (adapted from
Ref.~\cite{Servant:2002aq}).}
\label{fig:ued}
\end{figure}

In minimal UED models, the LKP is the KK partner of the electroweak
hypercharge gauge boson $B^0$, denoted by $B^{(1)}$
\cite{Cheng:2002iz}. The relic density of LKP dark matter lies in the
range favoured by WMAP data if $1/R \simeq {\cal{O}}{\mathrm(1~TeV)}$
\cite{Servant:2002aq}. The $\nu^{(1)}$ state, another possible
candidate, is already excluded by present limits from direct detection
experiments~\cite{Servant:2002hb}. In the early universe $B^{(1)}
B^{(1)}$ annihilation proceeds through the s-channel and is more
efficient than in the SUSY neutralino case. The cosmologically
interesting mass scale for the LKP is therefore significantly larger
than that for supersymmetric candidates (see Figure~\ref{fig:ued}),
although it still lies well within the anticipated LHC reach of $\sim$
1.5 TeV \cite{Cheng:2002ab}. Co-annihilation with the $e^{(1)}_R$
state is possible if its mass is almost degenerate with that of the
LKP however the UED co-annihilation cross-section is not sufficiently
large to reduce significantly the $B^{(1)}$ density. The net result is
that the dark matter density increases in the case of nearly
degenerate states, since the $B^{(1)}$ and the $e^{(1)}_R$ decouple at
about the same freeze-out temperature.

Kaluza-Klein dark matter also offers interesting prospects for direct
or indirect detection at space- and ground-based
experiments~\cite{Servant:2002hb,Cheng:2002ej}. Direct detection (see
Section~\ref{sec:direct} for the SUSY case) proceeds through KK quark
and Higgs exchange between relic LKPs and atomic nuclei and can give
sensitivity to LKP masses in excess of 1 TeV for tonne scale
experiments and favourably small values of the level one quark and
$B^{(1)}$ mass difference.

There are three main indirect signatures which can be sought. The
first consists of a peak of mono-energetic positrons which could be
produced in $B^{(1)} B^{(1)} \to e^+e^-$ annihilation. The AMS
experiment on the ISS should be able to distinguish positrons from
photons up to about 1~TeV and consequently could provide a means to
detect LKP dark matter for masses up to $\simeq$1000~GeV. A second
signature is represented by an excess of photons, with a continuum
spectrum, which could probe mass scales up to about 600~GeV. Finally
high energy neutrinos could be detectable at foreseen experiments,
with a sensitivity up to about 400~GeV.

Non-supersymmetric grand unified theories with warped extra dimensions
also introduce a stable Kaluza-Klein state, which could be a candidate
for cold dark matter. Such a state arises from imposing baryon number
symmetry which is spontaneously broken at the Planck
scale~\cite{Agashe:2004ci}. It has the quantum numbers of a
right-handed neutrino and cannot decay into SM particles. Depending on
the other parameters of the model, there is a vast range of masses for
this particle to produce the required dark matter density, from a few
tens of GeV up to about 1~TeV or higher. At the LHC, a possible
signature of such a scenario is provided by the production of the
next-to-lightest KK particles which would have a suffiently long
lifetime to travel through the detector volume and be recorded as
heavy stable particles.

The main feature of UED models, as well as SUSY models, which leads to
the existence of a plausible dark matter candidate, is the existence
of a conserved quantum number distinguishing the new states from those
appearing in the SM. Conservation of these quantum numbers in general
provides these models with additional attractive features
\cite{Cheng:2003ju}. This idea has been generalised to solve the
so-called ``little'' hierarchy problem suffered by many models
predicting new physics at the TeV scale \cite{Cheng:2003ju}. In doing
so a candidate dark matter particle is invariably produced. One test
case considered by the authors of Ref.~\cite{Cheng:2003ju} was that of
a little Higgs model, in which a $Z_2$ symmetry referred to as
``T-parity'' is imposed. The result is a dark matter candidate which is
either the new $B'$ gauge boson or a new $SU(2)_W$ singlet or triplet
scalar particle. The annihilation of the former to SM particles is not
chirally suppressed, unlike in {\it e.g.} SUSY models, and so indirect
detection signatures are expected to be enhanced. Singlet scalar
candidates have very small couplings to other states and must
therefore be of very low mass ($\lesssim$ 100~GeV) to give an
acceptable relic density \cite{Cheng:2003ju}. States with a
significant triplet component can interact with light gauge bosons,
providing a mechanism for their annihilation in the early universe,
and consequently their cosmologically allowed mass range is expected
to be higher ($\gtrsim$ 500~GeV) \cite{Birkedal-Hansen:2003mp}.

\section{Direct Detection of DM particles}

\label{sec:direct}

\subsection{Motivation}

\label{sec:direct1}

Although LHC experiments can provide supporting evidence for the
existence of cold dark matter particles, their limited sensitivity to
the lifetime of neutral particles (e.g. $\tau \lesssim 1$ ms for
$\chioi \ra \tilde{G} \gamma$ in GMSB SUSY models in ATLAS
\cite{atltdr-susy}) prevents them from providing conclusive
proof. Similarly it is very difficult for astroparticle experiments to
prove that any observed signal is due to a particular candidate, such
as the LSP, rather than some other exotic form of new physics ({\it
e.g.} UED). This means that in order to obtain a discovery of a
specific new dark matter candidate \emph{ consistent} signals will
need to be observed at \emph{ both} colliders such as the LHC \emph{
and} in astroparticle experiments.

There are two classes of astroparticle searches for cold dark matter;
indirect searches seeking evidence of dark matter annihilation
products such as neutrinos generated in the earth, sun or galactic
center, or gamma rays or anti-matter generated in the galactic halo
(see e.g. Ref.~\cite{Feng:2000zu} and references contained therein),
and direct searches sensitive to the elastic scattering of cold dark
matter particles from atomic nuclei in earth-bound detectors
\cite{Goodman:1984dc}. In this review we concentrate on the latter,
however it should be noted that the former can also provide very
strong constraints and indeed may already have provided first evidence
for particle dark matter in the form of an excess of positrons from
the galactic halo \cite{Coutu:1999ws}.

\subsection{Basics of Direct Detection}

\label{sec:direct2}

In direct searches, evidence for the existence of cold dark matter
particles is provided by an observation of an anomalous source of
recoiling nuclei inside a detector. The precise form of the recoil
energy spectrum of the nuclei depends on factors arising from
astrophysics (the local density and velocity distribution of incident
dark matter particles), nuclear physics (form-factors and coupling
enhancement factors for scattering at finite momentum transfer $q$)
and particle physics (the particle mass and interaction
cross-section). In general, the spectrum is expected to fall quickly
with increasing energy due to the small average kinetic energy of the
population of dark matter particles trapped within our galaxy (the
galactic ``halo''). This leads to a need for detectors with extremely
low ($\sim$ keV) nuclear recoil energy thresholds. In addition, the
nuclear scattering cross-section is strongly enhanced by a factor
$\propto A_m^2$ (nuclear target mass squared) for spin-independent
interactions (see below), favoring target materials incorporating high
mass (large $A_m$) nuclei.

The main background to signal events due to nuclear recoils in these
detectors arises from electron recoils generated by Compton scattering
of gamma radiation from naturally occurring radio-nuclides in, and
around, the target. It is therefore essential that detectors be
shielded from external sources of gamma radiation, and that they be
constructed from materials of the highest possible radio-purity. It is
also most advantageous if detectors are designed to discriminate
between nuclear recoil signal and electron recoil background events.

In addition to the electron recoil background there is also a
potential source of irreducible background events from elastic
scattering of neutrons from nuclear spallation by cosmic ray muons and
decay of radio-nuclides from natural U and Th chains in the detector
and its environment. The flux of spallation neutrons can be reduced to
very low levels by installing detectors deep underground where the
muon flux is minimal. Neutrons from radioactive decay are generally of
lower energy and can be absorbed using hydrogenous shielding materials
or tagged using active neutron vetoes.

\subsection{Direct Searches and SUSY}

\label{sec:direct3}

Direct searches for cold dark matter rely upon the existence of
weak-scale couplings between dark matter particles and quarks and
gluons in the nucleus. The nature of these couplings depends strongly
on the identity of the dark matter particle with scalar, axial-vector
and vector couplings being important for different candidates
\cite{Jungman:1995df}. Axial-vector currents couple to the
distribution of spin in the nucleus and hence contribute to the \emph{
spin-dependent} elastic scattering cross-section. Scalar currents by
contrast couple effectively to the mass distribution of the nucleus
and contribute to the \emph{ spin-independent} elastic scattering
cross-section. For non-Majorana dark matter particles vector couplings
tracing the charge distribution of valence quarks in the nucleus can
further contribute to the spin-independent cross-section
\cite{Jungman:1995df}. It is important to note that the
spin-independent cross-section is a measured quantity and is, in this
case, not equivalent to the scalar cross-section, which must be
calculated using model assumptions.

In gravity-mediated MSSM cold dark matter models the LSP can be the
lightest neutralino, a sneutrino, or possibly a light gluino (see
Section~\ref{sec:beyond}). As noted above, the last possibility is
heavily constrained and will not be discussed further. The sneutrino
has a large coupling to ordinary matter and hence is almost excluded
\cite{Falk:1994es} by direct searches due to a large predicted vector
coupling contribution to the spin-independent scattering
cross-section. This leaves the Majorana $\chioi$ as the most favored
MSSM cold dark matter candidate.

In CMSSM models when the LSP is neutral it is always the $\chioi$. At
tree-level the scalar $\chioi$-nucleus interaction receives
contributions from $\tilde{q}-\chioi$ couplings involving t-channel
Higgs exchange or s-channel $\tilde{q}$ exchange. At 1-loop further
significant contributions can arise from diagrams coupling the
$\chioi$ to a gluon \cite{Drees:1993bu}. Axial-vector interactions
arise from t-channel $Z^0$ exchange and s-channel $\tilde{q}$
exchange. The magnitudes of the scalar and axial-vector cross-sections
depend on a number of SUSY parameters including the $\chioi$ mass and
composition ({\it i.e.} $M_1$, $\tgbet$ and $\mu$), the squark masses
and the heavy Higgs mass $m_A$. In order to obtain a model-independent
estimate at the LHC of the expected event rates in direct search
experiments it will therefore be necessary to perform a wide range of
SUSY particle mass measurements.

Within the CMMSM, typical values for the scalar $\chioi$-nucleon
interaction cross-section range from $\sim$ $10^{-6}$ pb to $\sim$
$10^{-8}$ pb in the focus point region, where $\chioi \chioi A$
couplings are enhanced by the Higgsino content of the $\chioi$ and
$Aqq$ couplings by the large value of $\tgbet$ \cite{Feng:2000gh}, to
$\lesssim$ $10^{-10}$ pb in the co-annihilation tail
\cite{Baer:2003jb}. When $\mu<0$ it is possible for the $u$ and $d$
quark contributions to the scattering to interfere destructively
reducing the cross-section still further
\cite{Baer:2003jb,Ellis:2000ds}. Large values of the scalar
cross-section are also obtained in regions with small $\mo$ and
$\mhlf$ where the squark mass is small leading to efficient s-channel
scattering. Similar arguments apply to the axial-vector
$\chioi$-nucleon cross-section, with values ranging from $10^{-4}$ pb
in the focus point region, where the Higgsino content of the $\chioi$
enhances the $Z^0\chioi\chioi$ coupling, down to $10^{-8}$ pb in the
stau co-annihilation tail, where $m_{\sql}$ is large
\cite{Ellis:2000ds}.

\subsection{Prospects for Direct Searches}

\label{sec:direct4}

\begin{figure}
\begin{center}
\epsfig{file=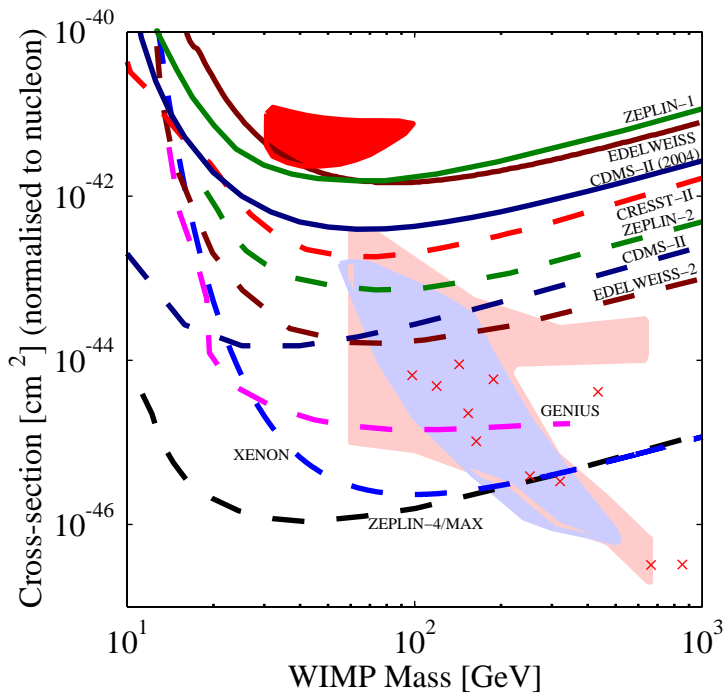,width=10cm} 
\end{center}
\caption{\label{dm-reach} Sensitivities of some running and planned
direct detection dark matter experiments to the spin-independent
elastic scattering cross-section. Full curves correspond to limits
from existing experiments (ZEPLIN-1 \cite{zeplin1}, EDELWEISS
\cite{Benoit:2001zu}, CDMS-II (2004)\cite{Akerib:2004fq}), dashed
curves to predicted sensitivities of future experiments (next
generation: CRESST-II \cite{Jochum:2000hi}, ZEPLIN-2
\cite{Luscher:yn}, CDMS-II \cite{Perera:hd} and EDELWEISS-2
\cite{Chardin:2003qp}; tonne-scale: GENIUS
\cite{Klapdor-Kleingrothaus:aw}, XENON \cite{Aprile:2002ef} and
ZEPLIN-4/MAX \cite{Spooner:2001hn,Cline:2003zx}). The full dark (red)
region corresponds to the 3 $\sigma$ allowed region from the DAMA
experiment \cite{Bernabei:2000qi}. The full light regions correspond
to predictions by Chattopadhyay et al. \cite{Chattopadhyay:2003qh} and
Baer et al. \cite{Baer:2003jb} in the light of WMAP data. The crosses
correspond to neutralino masses and cross-sections predicted for
post-LEP benchmark CMSSM models \cite{Battaglia:2003ab}. Adapted from
Ref.~\cite{dmplotter}.}
\end{figure}

The current generation of direct search experiments have attained a
sensitivity to the normalized spin-independent $\chioi$-nucleon cross
section $\sigma_{si} \sim$ $4\times 10^{-7}$ pb \cite{Akerib:2004fq}
(Figure~\ref{dm-reach}). The DAMA experiment operating at the Gran
Sasso underground laboratory \cite{Bernabei:1998td} has claimed
evidence for an annual modulation signature consistent with a cold
dark matter interpretation ($m_{\chioi} \sim 52$~GeV, $\sigma_{si}
\sim 7.2 \times 10^{-6}$ pb \cite{Bernabei:2000qi}), however the
region of parameter space allowed by this data has now been largely
excluded by other experiments
\cite{Akerib:2004fq,Abrams:2002nb,Benoit:2001zu,zeplin1}. The region
allowed by the DAMA data can be extended by varying the parameters of
the velocity distribution of cold dark matter particles trapped within
our galaxy \cite{Bernabei:2000qi}, or by assuming contributions to the
signal from both spin-dependent and spin-independent couplings
\cite{Bernabei:ve}. Nevertheless it is not clear that these new
regions of parameter space are not also excluded by other experiments
when similar assumptions are made. Current limits on the
spin-dependent $\chioi$-nucleon cross-section $\sigma_{sd}$ assuming a
pure Higgsino LSP candidate (see below) are $\sim$ 1 pb
\cite{Ahmed:2003su}.

Within the next two years a new generation of high mass experiments
capable of enhanced electron-recoil background discrimination will
come on-line. These are sensitive to spin-independent
interaction cross-sections $\sim$ $10^{-8}$ pb after a few years of
running \cite{Luscher:yn,Perera:hd,Chardin:2003qp,Jochum:2000hi}. In
the longer term, specific proposals are being made to build still
larger tonne-scale experiments
\cite{Schnee:2002jx,Spooner:2001hn,Cline:2003zx,Aprile:2002ef,Klapdor-Kleingrothaus:aw}
capable of probing $\sigma_{si} \sim 10^{-10}$ pb at the lower limit
of the CMSSM favored cross-section region (at least for
$\mu>0$). These experiments are likely to be particularly challenging
because of the difficulty of ensuring that neutron-induced background
events do not systematically limit detector sensitivity. It is likely
that the neutron flux from external sources can be reduced to
acceptable levels with active and passive shielding
\cite{Carson:2004cb}.  Reduction of the neutron flux from U/Th
contamination of detector components will require careful design and
material assay. This background will probably define the
ultimate sensitivity of direct search experiments since even the
purest detector construction materials will always contain some U/Th
chain nuclides \cite{Carson:2004cb}. Improvements in the sensitivity
to spin-dependent interactions can be expected to be similar to those
for spin-independent interactions, however the lack of a cross-section
enhancement due to nuclear coherence \cite{Jungman:1995df} means that
tonne-scale detectors will only be sensitive to $\sigma_{sd} \sim
10^{-5}$ pb, which is only just inside the CMSSM predicted range.

By 2011, when the LHC should have acquired  100 fb$^{-1}$ of
data, tonne-scale direct search experiments should have been running
for a few years. By this stage therefore cold dark matter models
predicting $\sigma_{si} \gtrsim$ $10^{-9}$ - $10^{-10}$ pb should have
either been excluded or produced a signal in these experiments. This
opens up the exciting prospect of complementary studies of particle
dark matter using direct search data and exclusive measurements at the
LHC.

\section{Complementarity of Measurements}

\label{sec:complement}

Direct search experiments are complementary to the LHC in several
important ways. In terms of discovery of SUSY it is clear that direct
searches are particularly sensitive to focus point models in which the
$\chioi$ acquires a significant Higgsino component. These are the same
models however which the LHC will find hardest to discover due to the
high masses and hence reduced production cross-sections of both
squarks and gluinos. As a result the LHC and direct search discovery
contours are roughly orthogonal in this region of parameter space and
the latter could conceivably find evidence for BSM physics at mass
scales in excess of 3 TeV with $\sigma_{si}$ as high as $10^{-8}$ pb
\cite{Baer:2003jb}. By contrast, in the co-annihilation tail the
sensitivity of direct search experiments drops off rapidly with rising
$\mhlf$ as $m_{\sql}$ increases, and here the LHC stands a much better
chance of making a discovery.

In the event of discovery of signals at both the LHC and direct
detection experiments further valuable cross-fertilization is
possible. If the LHC experiments can measure the SUSY model parameters
({\it e.g.} $\mo$, $\mhlf$ etc. in the CMSSM) using the techniques
described in Section~\ref{sec:lhc_excl} then a model-dependent
estimate of the spin-independent and spin-dependent elastic scattering
cross-sections can also be obtained. It may also be possible to
measure sufficient weak-scale SUSY parameters, including the
neutralino mass matrix parameters together with $m_A$ and $m_{\sql}$,
to obtain a ``model-independent'' estimate of the cross-sections for
comparison with direct search measurements. Note that the precision of
these measurements will itself be ultimately limited by astrophysical
model-dependence due to the unknown distribution of dark matter within
the galactic halo. It is important to remember however that the direct
search parameter space is defined both by the interaction
cross-section and the dark matter particle mass. The kinematics of the
elastic scattering process mean that for $\chioi$ masses $\lesssim$
$A_m$ (the mass of the target nucleus) the shape of the energy
spectrum of signal nuclear recoils for scalar interactions is strongly
correlated with the mass of the $\chioi$ (for larger $\chioi$ masses
the energy spectrum tends to a limiting distribution insensitive to
the mass value). This correlation is furthermore rather
model-independent. Consequently for $\chioi$ masses $\lesssim$ 180~GeV
($\sim$ the mass of a tungsten nucleus, the heaviest nucleus used in
current or proposed experiments \cite{Jochum:2000hi}) it should be
possible to measure the $\chioi$ mass rather accurately without
recourse to any particular model framework. This will then allow
direct comparison with LHC results.

It is interesting to note that a good case study of a direct search
mass measurement already exists in the form of the analysis performed
by DAMA of their claimed annual modulation signal
\cite{Bernabei:1998td,Bernabei:2000qi,Bernabei:ve}. The soft energy
spectrum observed in this experiment is claimed to be consistent with
a small $\chioi$ mass considerably less than that of the $^{127}$I
target nuclei. Under this assumption therefore an accurate measurement
of the $\chioi$ mass can be performed giving a precision $\sim$ 20 \%
($52^{+10}_{-8}$~GeV \cite{Bernabei:2000qi}).

LHC-direct search complementarity is likely to be even more pronounced
if a spin-dependent dark matter signal is observed. In this case,
there is a strong dependence of the relative magnitude of the
axial-vector couplings of the $\chioi$ composition to protons and
neutrons \cite{Tovey:2000mm} that can lead to significant
model-dependence in the overall $\chioi$-nucleus (or normalized
$\chioi$-nucleon) cross-section. Only in the pure Higgsino case is
this model-dependence lifted. Direct search experiments can absorb
this model-dependency into a redefinition of measured cross-sections
in terms of cross-sections for scattering from protons and neutrons
separately \cite{Tovey:2000mm}. Only with the aid of LHC measurements
of {\it e.g.} the neutralino mass matrix however can these results be
combined to obtain an overall experimental measurement of the
spin-dependent scattering cross-section. 

The variation of the relative magnitude of $\chioi$-p and $\chioi$-n
axial-vector couplings with $\chioi$ composition also leads to some
residual model dependency in the shape of the energy spectrum obtained
from spin-dependent scattering. The form-factor $S(q)$ relating
$\chioi$-nucleus scattering at finite momentum transfer $q$ to that at
zero momentum transfer contains terms depending on isoscalar and
isovector parameters $a_0$ and $a_1$ related to $\chioi$-p and
$\chioi$-n coupling parameters $a_p$ and $a_n$ \cite{Ressell:1997kx}:
\begin{equation}
S(q)=a_0^2S_{00}(q)+a_1^2S_{11}(q)+a_0a_1S_{01}(q)
\end{equation}
where 
\begin{equation}
a_0 =a_p+a_n, \bigskip a_1=a_p-a_n.
\end{equation}
Here the $S_{ij}(q)$ are independent constituent nuclear
form-factors. The differing $q$ dependence of these $S_{ij}(q)$
form-factors means that $S(q)$ differs for differing $a_p/a_n$ and
hence differing $\chioi$ composition. This leads to a SUSY model
dependency in the shape of the overall energy spectrum. This
model-dependency is generally rather small ($\lesssim$ 10 \% for low
energy recoils in e.g. $^{127}$I) however the accuracy of direct
search measurements of $\sigma_{sd}$ and $m_{\chioi}$ for comparison
with LHC predictions could conceivably be improved with an estimate
from LHC data of the neutralino mass matrix.

\section{Conclusions}

\label{sec:concl}

Recent astrophysical measurements from WMAP and other experiments have
strengthened the case for cold dark matter in the universe. Over the
next decade the hypothesis of particle cold dark matter will be
stringently tested in a number of different ways both with colliders
such as the LHC and with non-accelerator experiments such as direct
dark matter searches. If SUSY or other BSM physics is discovered at
the LHC then measurements of the properties of the new particles can
be used to predict the properties of dark matter signals in CMB and
astroparticle experiments. Similarly if dark matter is discovered in a
direct search experiment model-independent measurements of quantities
such as the mass of the dark matter particle can be compared with
signals observed at the LHC. Only by combining these results can we
confirm both the existence of non-baryonic cold dark matter and its
identity.

\section*{Acknowledgments}

The work of I.H. and M.B. was supported in part by the Director,
Office of Energy Research, Office of High Energy Physics of the
U.S. Department of Energy under Contract DE-AC03-76SF00098.
Accordingly, the U.S. Government retains a nonexclusive, royalty-free
license to publish or reproduce the published form of this
contribution, or allow others to do so, for U.S. Government
purposes. D.R.T. wishes to acknowledge PPARC for support.

\end{document}